\documentclass[12pt]{article}

\usepackage[mathscr]{eucal}
\usepackage{amsmath,amsfonts,amssymb,amsthm}
\usepackage[hypertex]{hyperref}
\usepackage[matrix,arrow]{xy}
\advance\textheight\topskip
\numberwithin{equation}{section} \makeatletter
\@addtoreset{equation}{section}

\theoremstyle{definition}

\newtheorem*{cohomology}{Theorem}

\newcommand{\bref}[1]{\textbf{\ref{#1}}}

\def\be{\begin{equation}}
\def\ee{\end{equation}}
\def\ba{\begin{array}}
\def\ea{\end{array}}
\def\d{\partial}
\def\dps{\displaystyle}

\newcommand{\half}{\frac{1}{2}}
\newcommand{\ads}{AdS_{2}}
\def\tr{{\rm Tr}}
\def\Hsalg{\textrm{hs}[\nu]}

\def\cA{\mathcal{A}}

\def\cG{\mathcal{G}}
\def\cH{\mathcal{H}}
\def\cI{\mathcal{I}}

\def\cM{\mathcal{M}}
\def\cN{\mathcal{N}}

\def\cP{\mathcal{P}}

\def\cR{\mathcal{R}}
\def\cS{\mathcal{S}}

\def\cU{\mathcal{U}}
\def\cV{\mathcal{V}}
\def\cW{\mathcal{W}}

\numberwithin{equation}{section}
\makeatletter
\@addtoreset{equation}{section}

\begin{document}

\renewcommand{\thefootnote}{$\dag$}

\begin{flushright}
FIAN-TD-2013-16 \\

\end{flushright}

\vspace{8mm}

\begin{center}

{\Large\textbf{Global and local properties of \\ \vspace{2mm}  $AdS_2$ higher spin gravity}}

\vspace{.9cm}

{\large K.B.     Alkalaev \footnote{email: alkalaev@lpi.ru}}

\vspace{0.5cm}

\textit{I.E. Tamm Department of Theoretical Physics, \\P.N.
Lebedev Physical Institute,
\\ Leninsky ave. 53, 119991 Moscow, Russia}

\vspace{1mm}
\textit{and}
\vspace{1mm}

\textit{Moscow Institute of Physics and Technology, \\
Dolgoprudnyi, 141700 Moscow region, Russia}
\vspace{0.5cm}

\begin{abstract}

Two-dimensional BF  theory with infinitely many  higher spin fields  is  proposed.
It is interpreted as the  $\ads$ higher spin gravity model  describing a consistent
interaction  between  local  fields  in $\ads$ space
including gravitational field, higher spin partially-massless fields,
and dilaton fields. We carry out analysis of  the frame-like and the metric-like formulation
of the theory.
Infinite-dimensional higher spin global algebras  and their finite-dimensional truncations
are realized in terms of $o(2,1) - sp(2)$ Howe dual auxiliary variables.

\end{abstract}

\vspace{5mm}
\end{center}

\newpage

{
\footnotesize
\tableofcontents
}

\newpage

\section{Introduction}

\renewcommand{\thefootnote}{\arabic{footnote}}
\setcounter{footnote}{0}

In the recent years, higher spin gauge theories  in three, four
and higher  dimensions have attracted considerable interest
(\textit{e.g.}, see reviews
\cite{Vasiliev:1999ba,Bekaert:2005vh,Giombi:2012ms,Gaberdiel:2012uj,Didenko:2014dwa}
and references therein),  while comparatively little attention has
been paid to two-dimensional higher spin theories
\cite{Bengtsson:1986zm,Fradkin:1989uh,Vasiliev:1995sv,Rey,Alkalaev:2013fsa,Grumiller:2013swa}.
One of the reasons for this is that higher spin gravity in two
dimensions  does not necessarily share some of characteristic
features of its higher dimensional cousins such as $(A)dS$  background geometry  or infinitely many propagating
massless modes of all spins.  So for example conventional $2d$
Fronsdal-type equations of motion both for  massless or massive
fields of higher spins  $s\geq 1$ do not propagate  local degrees
of freedom. For that matter the two-dimensional case is somewhat
analogous to that in three dimensions, where higher spin
Chern-Simons theory also describes  no local degrees of freedom
\cite{Blencowe1, Blencowe2,Henneaux:2010xg,Campoleoni:2010zq}.

It follows that in two dimensions the notion of higher spin gauge  fields  should be clearly defined.
We can, at least formally, introduce  gauge fields of higher ranks and impose one or another set of
gauge invariant equations and/or constraints. Then some of the resulting  gauge systems have
no local degrees of freedom, while others
describe matter  modes  as particular components of higher rank gauge fields. In the former case
the respective  gauge fields often result   from  higher dimensional gauge systems by taking $d=2$. In particular,
both global  and gauge transformations  remain intact, while local degrees of freedom disappear.

In view of the above  we propose to consider a particular $2d$ topological field theory as higher
spin  gravity with the cosmological constant. The theory is
formulated  as two-dimensional BF model  with $\cA$-valued
$0$-form and $1$-form fields, where $\cA$ is some finite-dimensional  or infinite-dimensional
higher spin Lie algebra. \footnote{Two-dimensional topological gravity
and its higher spin extensions can be defined in a different way as
topological field theories of Witten type  \cite{Labastida:1988zb}. The higher spin gravity
we elaborate here is obviously a topological field theory of Schwarz type.}
In \cite{Alkalaev:2013fsa} we explicitly considered
the finite-dimensional case of  $\cA = sl(N, \mathbb{R})$ for $N\geq 2$. The point is that the gauge algebra
can be  represented in the higher spin basis where  generators are arranged as subalgebra $sl(2, \mathbb{R})$ rank-$s$ irreps
so that the respective connections are identified with two-dimensional spin-$(s+1)$ fields. The case of $N=2$
corresponds to the Jackiw-Teitelboim dilaton gravity
\cite{Barbashov:1980bm,Fukuyama:1985gg}, while taking $N \geq 3$  gives rise to particular higher spin extensions.
The $N=3$ theory was also discussed  in
\cite{Grumiller:2013swa} in the framework of Poisson sigma-models, mainly form the holographic perspective.

It is remarkable that a  ground state of the model under consideration is
given by the $\ads$ spacetime.  It follows that the gauge sector of the $sl(N, \mathbb{R})$ higher spin gravity model
comprises   gauge   fields in $\ads$
space with spins $s = 2,3, ... , N$ and masses $m^2_s =
s(s-1)\Lambda$, where $\Lambda$ is the cosmological constant. Using their global symmetry properties one finds that
the fields are to be treated as  "topological  partially-massless" fields
of  maximal depth  \cite{Alkalaev:2013fsa}. Recall that  the system does not have local degrees of
freedom.  It follows that the $\ads$ higher spin gravity can be interpreted as a consistent theory of topological
yet interacting  partially-massless higher spin  fields given in
a closed form. It is worth noting that partially-massless fields in higher dimensions
do have local degrees of freedom \cite{Deser:1983tm,Deser:2001us,Zinoviev:2001dt,Skvortsov:2006at},
while their interactions at the action level are known only in the cubic
approximation \cite{Zinoviev:2006im}.

In this paper we formulate  $\ads$ higher spin gravity with  (in)finitely many fields as
BF theory  for  the infinite-dimensional  higher spin gauge algebra $\cA = \Hsalg$
and its finite-dimensional truncations \cite{Feigin,Vasiliev:1989qh}. Note that similar  models with an infinite higher spin algebra  were partly
discussed in \cite{Fradkin:1989uh, Rey}.  Here  we focus on the following issues.

\begin{itemize}

\item Local tensor fields in the $\ads$ higher spin gravity: frame-like versus the metric-like formulation. We study in detail
the interplay between the BF formulation of the higher spin gravity which
is actually the frame-like formulation and its  metric-like formulation which extends the original
Jackiw-Teitelboim dilaton gravity.

\item Global higher spin symmetry algebras: \footnote{By global symmetry algebra in topological field theory
we understand (generalized) Killing symmetries of a given vacuum solution to the theory. In a theory with local degrees of freedom this notion naturally
extends to conventional global symmetry algebras acting on the space of one-particle states.} a formulation using the Howe duality $o(2,1) - sp(2)$ between
$\ads$ global symmetry algebra and auxiliary  symplectic algebra.
We explicitly describe  previously unknown realization of  higher spin algebras $\cA = \Hsalg$ in terms of
$o(2,1)-sp(2)$ vector doublet variables. \footnote{The present construction of $\Hsalg$ uses
six independent oscillators which is a minimal number of variables allowing for the Howe duality. Other  approaches  with less number
of oscillators were known in the earlier literature
\cite{Pope:1989sr,Joung:2014qya}. } Gauging algebra $\cA$ defines local invariance  of the BF theory under consideration.

\item BF action for $\cA$-valued gauge fields: introducing particular trace operation on the infinite-dimensional gauge algebra
$\cA$ we define
various (in)finite-dimensional truncations directly at the action level. We study a perturbative expansion of the
action around the $\ads$ background.

\end{itemize}

\vspace{-2mm}

\noindent The outline of the  paper is as follows.

Section \bref{sec:lindyn}: The linearized  $\ads$ higher spin
gravity is formulated via  the BF action functional. The
action, the equations of motion, and the gauge symmetry
transformations are given explicitly. The BF formulation
under consideration is treated as a particular frame-like
formulation which is known to be a generalization of  the zweibein
description  of $2d$ gravitational systems. As a by-product, we propose a  higher spin generalization
of $2d$ Maxwell theory obtained  as higher spin BF theory extended by a particular quadratic potential.

Section \bref{sec:cohview}: BF systems are treated in the framework of the unfolded formulation
that pursues the cohomological understanding of  both lower spin and higher spin
systems (see the review \cite{Bekaert:2005vh} for details). The section contains a
detailed discussion of various mathematical structures underlying
the cohomological interpretation of the dynamics. The main objects here
are  the so-termed  $\sigma_+$ and $\sigma_-$ nilpotent operators
acting on  the field  space of the model. Elements of the space are
differential $p$-forms taking values in
any rank $o(2,1)$ finite-dimensional irreps. Using the
$\sigma_{\pm}$-cohomology we perform a cohomological  reduction of
the initial field space to a certain  subspace: a transition  from the
frame-like formulation of the model to its metric-like form. We  compute
$\sigma_{\pm}$-cohomology groups that completely identify the local
structure of the (linearized) metric-like theory: gauge symmetry,
independent metric-like fields, equations of motion and their
Bianchi identities.

Sections \bref{sec:offshell} and \bref{sec:0forms}:
Nilpotent operators $\sigma_{+}$ and $\sigma_-$ correspond to two different
cohomological reductions of the initial field space. So, in the
one-form sector of the BF higher spin model we find that the
system is equivalent either to  massive scalar theory with a mass
proportional to the cosmological constant and dependent on the
spin, or to higher rank current conservation conditions. The
scalar/current equations are invariant with respect to particular
type of trivial on-shell symmetries/improvements that eliminate all local degrees of freedom.   We suggest
that these two forms of a single system are analogous to the
well-known classical duality phenomenon occurring in the WZNW
theory when second-order equations can be represented as the
first-order conservation condition
\cite{Nappi:1979ig}. The same analysis is done in
the zero-form sector of the model.

Section \bref{Sec:summaryMETR}: It summarizes the metric-like
formulation developed in the previous sections. We list  the
metric-like equations of motions in the zero-form and one-form
sectors of the BF higher spin gravity model in both cases of
the $\sigma_{\pm}$ cohomological reductions. Finally, the model is interpreted as the higher spin
gauge-dilaton theory extending  the Jackiw-Teitelboim dilaton gravity. Also, we consider two metric-like
action functionals which give rise to dual metric-like equations
of motion. We find out that the BF action is a "parent" action
for the two dual metric-like formulations.

Section \bref{sec:higher spin algebra}:  Using  manifestly
covariant $o(2,1)-sp(2)$ vector notation we elaborate a
realization of the one-parametric  higher spin algebra $\Hsalg$ introduced in Refs.
\cite{Feigin,Vasiliev:1989qh}. Our realization is derived from the general
$d$-dimensional oscillator description  of the Eastwood-Vasiliev higher spin
algebra for $d\geq 3$ \cite{Eastwood:2002su,Vasiliev:2003ev}. The approach is
based on the Howe dual pair $o(2,d-1)-sp(2)$ realization in the
bimodule of formal power series in auxiliary doublet  variables
\cite{Vasiliev:2003ev,Vasiliev:2004cm}. Specifying to $d=2$ we find out that $\Hsalg$ is to be identified as
quotient algebra
obtained by singling out a particular ideal. The Howe duality
 $o(2,1) - sp(2)$ used to  describe  quotient higher spin algebras
may be useful in many respects, in particular, for considering general
non-linear two-dimensional higher spin models not necessarily of BF type. Indeed, the Howe
duality is known to be crucial to built a consistent interacting
higher spin theory in $d\geq 4$ dimensions \cite{Vasiliev:2003ev}.

Section \bref{sec:action}: It defines the full non-linear BF formulation
of the $\ads$ higher spin gravity. Since the gauge algebras are realized
as quotient algebras, the corresponding BF actions are formulated using
particular projecting technique that allows to factor out elements of ideals
directly  inside the action.
Quadratic higher spin actions studied in section
\bref{sec:lindyn}  result from a linearization around the $\ads$ background solution.

Section \bref{sec:concl}: It summarizes our results and discusses future  research directions. Details of the $\sigma_{\pm}$-cohomology
computation are given in Appendix \bref{sec:appendixA}. Details of the projecting technique
are given in Appendix \bref{sec:appendixB}.

\section{Quadratic  higher spin BF action}
\label{sec:lindyn}

Let $\cG_s$ be a linear space of differential $p$-forms on a two-dimensional manifold
taking values in finite-dimensional $o(2,1)$ totally symmetric and traceless representations of
arbitrary rank \footnote{A spacetime $\cM^2$ is a general two-dimensional manifold
with local coordinates $x^m$,  Lorentz world indices run $m,n = 0,1$, Lorentz fiber
indices run  $a,b = 0,1$, $o(2,1)$ fiber  indices  run
$A,B,C = 0,1,2$, $o(2,1)$ invariant metric is $\eta^{AB} = (+--)$.
The spacetime derivative is denoted as $\d_m = \d/\d x^m$, the de Rham differential
is $d = dx^m \d_m$.
The Levi-Civita tensor $\epsilon_{ABC}$ is normalized as $\epsilon_{012}=+1$.
Two-dimensional anti-de Sitter spacetime $AdS_2$ has a radius $L$
and a signature $(+-)$, so that the cosmological constant is
$\Lambda = -1/L^2$. The Levi-Civita tensor $\epsilon_{mn}$ is normalized as $\epsilon_{01} = +1$.
Symmetrization of indices has a unit weight and is labelled by parentheses.}
\be
\label{hsf}
F_{(p)}^{A_1 ... \, A_{s-1}} = dx^{m_1} \wedge  \cdots \wedge dx^{m_p}\,
\,F_{m_1 ...m_p}^{(A_1 ... \, A_{s-1})}\;,
\qquad
\eta_{BC}F_{(p)}^{BC A_3 ... \, A_{s-1}} =0 \;,
\ee
where $p=0,1,2$ is a rank of a differential form (at $p\geq 3$ differential  forms are
identically zero).
Using $o(2,1)$ Levi-Civita tensor
one shows that all non-symmetric finite-dimensional $o(2,1)$ irreducible representations either vanish identically,
or are described by hook-type traceless tensors
\be
\label{Hodge}
F_{(p)}^{A_1 ... \, A_{m}}\sim F_{(p)}^{A_1 ... \, A_{m},\, B_1 }\;.
\ee

Two-dimensional higher spin fields are defined to be elements of $\cG_s$. In two spacetime dimensions both massless and massive
Wigner groups trivialize and whence it follows that only scalar and spinor modes may propagate.
However, by a slight abuse of notation, we identify parameter $s$ as a spin.

When considering gravitational systems parameterized by the negative cosmological constant $\Lambda$, it is convenient
to represent gravitational fields as $o(2,1)$ connection $1$-forms
$W^A(x)\, T_A =  d x^m\, W^A_m(x)\, T_A$, where $T_A$ are $o(2,1)$ basis elements (see, \textit{e.g.},
\cite{Fukuyama:1985gg}).
Using antisymmetric basis one represents  the connection
as $W_m^{AB} = -W_m^{BA}$ which is dual to the original connection via
 $W_{m}{}_{AB} = \epsilon_{ABC}\,W_{m}^C$.
Flat connections satisfy the zero-curvature condition, which component form is
given by
\be
\label{zerocurv}
\cR^A_{mn} \equiv \d_mW^A_n - \d_n W^A_m   -  \epsilon^{ABC}\,W_{m,\,B}\,W_{n,\,C} = 0\;.
\ee

The frame field and Lorentz spin connection are introduced in a standard fashion using
the compensator $V^A$ normalized such that $V^AV_A = -L^2$.
In what follows, we use $V^A$ in the form $V^A = (0,0,L)$. The $o(2,1)$ covariant
decomposition of $W^A_m$ is given by
\be
W^A_m = E^A_m +  V^A \omega_m \;,
\ee
where the transversality conditions $V_A E^A_m=0$ and $\omega_m  = \Lambda\, W_m^A V_A$
give rise to $E_m^A = (e_m^a,0)$ and
$W_m^A = \big(\,e^a_m, -1/\sqrt{-\Lambda}\,\omega_m\big)$.

It is well-known that $AdS_2$ spacetime solves  constraint
\eqref{zerocurv}. The corresponding connection will be  denoted $W_0 = \big(h^a_m, -1/\sqrt{-\Lambda}\, w_m\big)$.
The zero-curvature constraint expresses Lorentz spin connection $w_m$
via the frame $h^a_m$, while the latter defines $AdS_2$ spacetime metric $g_{mn}$ through the standard identification
$g_{mn} = \eta_{ab} h^a_m h^b_n$, where $\eta_{ab} = (+-)$ is the fiber Minkowski metric.

%

Let us consider particular elements of the space $\cG_s$ which are  $0$-form field
$\Phi^{A_1 ... A_{s-1}}$, $1$-form field $\Omega^{A_1 ... A_{s-1}}$ along with
$2$-form field strength
\be
\label{gaugefield}
\Phi^{A_1 ... A_{s-1}}\,,
\qquad
\Omega^{A_1 ... \, A_{s-1}} = dx^{m}\, \Omega_{m}{}^{A_1 ... \, A_{s-1}}\,,
\qquad
R_1^{A_1 ... \, A_{s-1}}  = D_0\Omega^{A_1 ... \, A_{s-1}}\;,
\ee
where $D_0$ is  $o(2,1)$ covariant background derivative,
\be
\label{D0}
D_0 F_{(p)}^{A_1 ... A_k} = d T_{(p)}^{A_1 ... A_k} + \epsilon^{BC (A_1 }\,W_0{}_{B} F_{(p)}{}_C{}^{A_2 ... A_k)} + ... +
\epsilon^{BC (A_k }W_0{}_{B} T_{(p)}{}_C{}^{A_1 ... A_{k-1})}\;.
\ee
From now on, we systematically omit the wedge  product symbol $\wedge$. Representing  the zero-curvature condition
\eqref{zerocurv} evaluated on the background connection $W_0$ as $\cR(W_0) \equiv D_0 D_0 = 0$ one observes
that  higher spin field strengths are invariant with respect to the following gauge transformations
\be
\label{transOmega}
\delta \Omega^{A_1 ... \, A_{s-1}} = D_0 \xi^{A_1 ... \, A_{s-1}}\;,
\ee
where gauge parameters $\xi^{A_1 ... \, A_{s-1}}$ are $0$-forms taking values in the same finite-dimensional
representations. Note that the Bianchi identities $D_0 R$ in two dimensions are trivial since any $3$-form
vanishes identically. The 0-form fields are assumed to be  gauge invariant,
\be
\label{transPhi}
\delta \Phi^{A_1 ... A_{s-1}} = 0\;.
\ee
Fields \eqref{gaugefield} are referred to as frame-like fields as these generalize the gravitational connection
$W^A_m$ to any number of fiber indices and any rank of differential form.

Let us consider now  the BF action for a single rank-$s$ system,
\be
\label{actBFlin}
S_0[\Omega, \Phi] = \int_{\cM^2}  \Phi_{A_1 ... A_{s-1}}R_1^{A_1 ... A_s-1}\;.
\ee
The equations of motion obtained by varying   with respect to  $\Phi_{A_1 ... A_{s-1}}$
and $\Omega_{A_1 ... A_{s-1}}$ are
\be
\label{linBFeom}
\ba{l}
\dps
 R_1^{A_1 ... A_{s-1}} = 0\;,
\qquad
D_0 \Phi^{A_1 ... A_{s-1}} = 0\;.
\ea
\ee
Both the action and the equations of motion are invariant with respect to gauge transformations \eqref{transOmega} and \eqref{transPhi}.
In Section \bref{sec:action} the action \eqref{actBFlin} will be  obtained from a full non-linear BF
higher spin action by a linearization around $\ads$ background  $W_0$.

The original BF theory \eqref{actBFlin} can be deformed   in various ways. For instance, augmenting its action by a quadratic term
\be
\label{actBFlinAUGM}
S_0[\Omega, \Phi] = \int_{\cM^2} \Big( \Phi_{A_1 ... A_{s-1}}R_1^{A_1 ... A_s-1} -  \half\, \Phi^{A_1 ... A_{s-1}}\Phi_{A_1 ... A_{s-1}}\cV_2\Big) \;,
\ee
where $\cV_{2} = \epsilon_{ab}\,h^a \wedge h^b$ is the volume $2$-form built of $\ads$ background
frame fields, one obtains the following equations
\be
\label{augment}
 R_1^{A_1 ... A_{s-1}} = \cV_2\Phi^{A_1 ... A_{s-1}}\;,
\qquad
D_0 \Phi^{A_1 ... A_{s-1}} = 0\;.
\ee
Eliminating the auxiliary field $\Phi^{A_1 ... A_{s-1}}$ by  using its own equation of motion
one arrives at the action of the form
\be
\label{214}
S_0[\Omega] = \int_{\cM^2} R^\star_1{}_{\,A_1 ... A_{s-1}}R_1^{A_1 ... A_s-1}\;,
\ee
where $R_1^\star$ is the Hodge dual field strength. Note that now the action explicitly depends on the
background $\ads$ metric.
The rank-$s$ equations of motion following from \eqref{214}
\be
D_0^m R^{A_1 ... A_{s-1}}_{1\, mn } = 0\;,
\ee
generalize the Maxwell equations and describe no local degrees of freedom
(see also  our comments in the end of Section \bref{sec:weyl}).
For the simplest case
$s=1$ the action \eqref{actBFlinAUGM} is the well-known action for the Maxwell field $A_m$ on the background
metric $g_{mn}$ with the auxiliary scalar variable $f$: $\dps S_0[A, f] = \int d^2 x \sqrt{g}\,\big( f \epsilon_{mn} F^{mn} - \half f^2 )$,
where $F_{mn} = \nabla_mA_n - \nabla_n A_m$.
Representing the Maxwell action in this form is useful in the analysis of $2d$ Maxwell-dilaton theories
of gravity, since the dynamical field enters the action linearly (see,  \textit{e.g.}, \cite{Hartman:2008dq}).

\section{Cohomological view of BF equations}
\label{sec:cohview}

In order to analyze the dynamical content  of the BF action \eqref{actBFlin}
we employ homological tools developed within the unfolded formulation
(see the review \cite{Bekaert:2005vh} for details).
Indeed, one observes that the BF  equations of motion
are explicitly formulated as  the zero-curvature and the covariant constancy
conditions imposed on the frame-like fields which are differential forms taking values in certain
$o(2,1)$ irreps, see  \eqref{linBFeom}. Fortunately, such a geometrical setting naturally fits
the unfolded formulation.

Most importantly, using the unfolded machinery helps to obtain  the
metric-like formulation of the BF theory. For instance, in order to obtain the Jackiw-Teitelboim dilaton gravity theory
from $o(2,1)$ BF theory one should carefully identify the metric and scalar fields along with auxiliary fields,
use local Lorentz symmetry to set an antisymmetric part of the zweibein equal to zero, split all the equations
into dynamical and constraint ones \cite{Fukuyama:1985gg}. It is remarkable that  all
these operations can be  done in a systematic manner using
cohomology groups of certain nilpotent operators called  $\sigma_{\pm}$ acting on the field
space $\cG_{s}$ \eqref{hsf}. In other words, using the  $\sigma_{\pm}$-cohomology provides precise guidelines how to
pass from a frame-like (\textit{i.e.}, BF) formulation
to a metric-like formulation where the higher spin fields are higher rank Lorentz tensor fields.

In order to make using the cohomological methods more manifest  it is convenient to reformulate given BF equations
as off-shell system. It means that the right-hand-sides of BF equations are not zero
but some arbitrary sources. Sending the sources to zero implies going on-shell.
Indeed, put  equations \eqref{linBFeom}  off-shell as follows
\be
\label{2eq}
D_0 \Phi^{A_1 ... A_{s-1}} = B_{(1)}^{A_1 ... A_{s-1}}\;,
\ee
\be
\label{1eq}
D_0 \Omega^{A_1 ... A_{s-1}} = C_{(2)}^{A_1 ... A_{s-1}}\;,
\ee
where the right-hand-sides of the equations are now arbitrary differential $1$-form and $2$-form, respectively,
taking values in rank-$(s-1)$ irreducible $o(2,1)$ representation, \eqref{hsf}. By definition, sources $B_{(1)}$ are $C_{(2)}$
are invariant  with respect to gauge symmetry transformations \eqref{transOmega} and \eqref{transPhi}, and therefore
the off-shell system \eqref{2eq}-\eqref{1eq} retains the same gauge symmetry as the on-shell  one
\eqref{linBFeom}.

\subsection{$\sigma_{\pm}$ operators}
\label{sec:sigmas}

Most conveniently, the cohomological analysis of off-shell $o(2,1)$ covariant equation system  \eqref{2eq} - \eqref{1eq} is performed
in terms of Lorentz $o(1,1)\subset  o(2,1)$ algebra component fields. To this end, we rewrite elements of the
field space $\cG_s$ \eqref{hsf} in Lorentz basis,
\be
\label{dectraceless}
T_{(p)}{}^{A_1 ... \, A_{s-1}} = \bigoplus _{k=0}^{s-1}\; T_{(p)}{}^{a_1 ... \, a_k}\;,
\ee
where Lorentz fields are totally symmetric and traceless,
\be
\label{symprop}
T_{(p)}{}^{a_1 ... \, a_k} = T_{(p)}{}^{(a_1 ... \, a_k)}\;,
\qquad
\eta_{bc}\, T_{(p)}{}^{bca_3 ... \, a_k} = 0\;.
\ee
Therefore, in Lorentz basis space  $\cG_s$ is given by a direct sum of subspaces spanned by
differential $p$-forms $T^{a_1 ... a_k}_{(p)}$  with fixed value of $ k = 0, ...\,,  s-1$.
Such elements will be denoted as $T_{(p)}(k)$.
It is worth recalling   that a $o(1,1)$ totally symmetric and traceless  tensor  $T^{a_1 ... a_k}$
has just two independent components. This is most obvious in the light-cone parametrization
 $T^{a_1 ... a_k} \sim (T^{++ \cdots +},\, T^{-- \cdots -})$, where a number of $\pm$ equals $k$.
However, keeping  $o(1,1)$ symmetry manifest is convenient when analyzing the
dynamical content of the theory.

The space $\cG_s$ incorporates all  tensor fields of the theory,
including zero-forms, one-forms and associated two-forms
\eqref{gaugefield}, along with their $0$-form gauge symmetry
parameters   \eqref{transOmega}. For a given spin $s$  there are two natural gradings
in the space $\cG_s\,$: by a rank of differential
forms and by a number of Lorentz indices. On the other hand, there exist two
nilpotent algebraic operators acting on $\cG_s$ that shift the gradings by one.

Let us define   operators  $\sigma_\pm$ acting on $\cG_s$ as
follows $\sigma_\mp: \; T_{(p)}(k\pm 1) \rightarrow T_{(p+1)}(k)$.
Their component action is given by
\footnote{\label{f4}It stands to mention that  conventional $\sigma_-$ operator in $d\geq 4$ dimensions
turns to  $\sigma_+$ in $d=2$ dimensions. This is because in the case $d\geq 4$  the field space $\cG_s$ consists
of two-row rectangle
$o(2,d-1)$ traceless tensors that are replaced by one-row $o(2,1)$ traceless tensors in the case of $d=2$. In the
spin-$2$ case
this is achieved by using the Levi-Civita tensor
what changes the roles of $\sigma_-$ and $\sigma_+$ operators in dualized pictures. Note, however,  that this
difference is purely notational.}
\be
\label{sigmas}
\ba{l}
\dps
\sigma_-:\quad\alpha_{(k)} \, h_c\, T^{ca_1 ... a_k }_{(p)} = T^{a_1 ... a_k}_{(p+1)}\;,
\\
\\
\dps
\sigma_+:\quad \beta_{(k)}\Big[ h^{(a_1} \, T^{a_2 ... a_{k})}_{(p)}
+ \gamma_{(k)}\,\eta^{(a_1a_2} h_c\, T_{(p)}^{ca_3 ... a_k)} \Big]= {T}_{(p+1)}^{a_1 ... a_k}\;,
\ea
\ee
where $h^a_{m}$ is the $\ads$ background frame, while exact expressions for coefficients $\alpha_{(k)}, \beta_{(k)}$ and $\gamma_{(k)}$ are
given below, see \eqref{sigmakoef} and \eqref{gammakoef}. The operators satisfy
\be
\label{sigmascond}
\sigma_{-}^2=0\;,
\qquad
\sigma_+^2 = 0\;, \qquad \nabla^2 + \sigma_- \sigma_++ \sigma_+ \sigma_- =0\;,
\ee
where  covariant derivative $\nabla_{m} = \d_m + w_m$ is evaluated  with respect to $\ads$ background
Lorentz spin connection $w_m$. It is worth noting that conditions \eqref{sigmascond}
can be understood as realization of the zero-curvature condition $D_0^2 =0$ \eqref{zerocurv} in the Lorentz component
basis \cite{Vasiliev:2001wa},
\be
\label{derdec}
D_0 = \nabla + \sigma_- + \sigma_+ \;.
\ee

It is convenient to define the Euler operator $N$ counting a number of Lorentz indices,
$N T_{(p)}(k) = k T_{(p)}(k)$. Then, $[N, \sigma_{\pm}] = \pm \sigma_{\pm}$ and $[N,\nabla] = 0$.
Operator $N$ provides the space  $\cG_s$ with a finite grading,
\be
\label{grading}
\cG_s = \cG_s^{(0)} \oplus  \cdots \oplus \cG_s^{(s-1)} \;,
\ee
where a subspace  $\cG_s^{(k)}$ is spanned by homogeneous elements of degree  $k$. By definition, operator $\sigma_-$
decreases a  degree by one, operator $\sigma_+$ increases a degree by one.

The space $\cG_s$ can be endowed with an inner product given by
\be
\label{inner}
\left< A| B \right> = \delta_{k,\,l} \delta_{m+n,\, 2} \int_{\cM^2}
A_{(m)}{}^{a_1 ... a_k}\, B_{(n)}{}_{a_1 ... a_l}\;,
\qquad A, B \in \cG_s\;.
\ee
Modulo an overall coefficient, operators  $\sigma_-$ and $\sigma_+$ are
mutually conjugated with respect to the above inner product. The following
properties are elementary:
\be
\label{pro1}
\quad \;\;\;\;\left< A | B\right>   = 0\;,
\qquad \qquad
A \in \cG_s^{(k)}\;,
\quad
B\in \cG_s^{(l)}\;,
\quad k \neq l\;,
\ee
\be
\label{pro2}
\left<\sigma_{\pm} A | B\right>  = 0\;,
\qquad\qquad \forall A \in \cG_s\;, \quad \forall B \in Ker\, \sigma_{\mp}\;.
\ee

\paragraph{Exact expressions for the coefficients.} Coefficients
$\gamma_{(k)}$ in \eqref{sigmas} are fixed by the algebraic symmetry conditions \eqref{symprop} as
\be
\label{gammakoef}
\gamma_{(1)} = 0\;,
\qquad
\gamma_{(k)} =  - \frac{1}{k-1}\;,
\qquad
k = 2,3 ...,  s-1\;.
\ee
Coefficients $\alpha_{(k)}$ and $\beta_{(k)}$ are defined by conditions \eqref{sigmascond}.
Namely, one arrives at the equation system,
\be
\rho_{(k)} \equiv  \alpha_{(k)}\beta_{(k+1)}\;: \qquad \Lambda + \rho_{(k)}\big[\gamma_{(k+1)}-1\big]+ \rho_{(k-1)} = 0\;,
\ee
for  $k = 1, ... \,, s-1$. The explicit solution reads
\be
\label{sigmakoef}
\rho_{(k)} = -\Lambda \frac{(s-k-1)(s+k)}{2(k+1)} \;.
\ee
Using proper  field redefinitions one
can set either $\beta_{(k)}=1$  or
$\alpha_{(k)} = 1$ for $ k =1, ...\,, s-1$  so that the solution is unique.  Here, we choose the
former case indicating that the dynamical systems under consideration are extended from Minkowski to AdS space.

\subsection{Cohomological analysis}
\label{sec:cohanal}

Below we shortly describe the general idea of  the cohomological reduction of the off-shell BF
system \eqref{2eq}-\eqref{1eq} using  $\sigma_{\pm}$ nilpotent operators
(see Ref.   \cite{Bekaert:2005vh} for more details).

Let us consider $p$-form gauge fields $\pi_{(p)}(k) \in \cG_s$. Then, using the decomposition \eqref{derdec},
the unfolded equations \eqref{2eq}, \eqref{1eq}  can be represented
in the Lorentz  component form  as follows
\be
\label{skelet}
\nabla \pi_{(p)}(k) + \sigma_- \pi_{(p)}(k+1) + \sigma_+ \pi_{(p)}(k-1) = Z_{(p+1)}(k)\;,
\ee
where differential $(p+1)$-forms $Z_{(p+1)}(k)$ are the sources, while $k = 0, ..., s-1$,
and  a rank of differential forms  runs $p=0,1$ since for $p=2$ the above expression
vanishes identically.

The unfolded equations \eqref{skelet} are invariant with respect to  gauge transformations given by
\be
\label{noch}
\delta \pi_{(p)}(k) = \nabla \varepsilon_{(p-1)}(k) + \sigma_- \varepsilon_{(p-1)}(k+1) + \sigma_+ \varepsilon_{(p-1)}(k-1)\;,
\ee
where $(p-1)$-forms $\varepsilon_{(p-1)}(k)$ are gauge parameters. In fact,
the gauge symmetry transformation appears at $p=1$ only. Indeed,  in the  case $p=0$ the  gauge fields have no associated gauge parameters,
while in the  case $p=2$ the corresponding equations of motion vanish identically.

Quantities   $Z_{(p+1)}(k)$ on the right-hand-side of \eqref{skelet} are not completely arbitrary
and are restricted by the Bianchi identity
\be
\label{bian}
\nabla Z_{(p+1)}(k) + \sigma_- Z_{(p+1)}(k+1) + \sigma_+ Z_{(p+1)}(k-1) = 0\;,
\ee
which is a differential $(p+2)$-form. It is obtained by using  conditions \eqref{sigmascond}.
For  $p=1$ the Bianchi identity is a $3$-form that  vanishes identically.

Note that the unfolded equations, gauge transformations and identities are decomposed
according to the grade degree \eqref{grading}. On the other hand, operators $\sigma_{\pm}$ enter
all equations algebraically. It suggests that the gauge system \eqref{skelet} -  \eqref{bian} can be analyzed
recurrently, starting either from the minimal grade degree $k=0$ equations or, from the maximal grade degree
 $k=s-1$ equations. In both cases, one arrives at  the linear systems of the type
\be
\label{sigmaeq}
\sigma_{\pm} X = Y\;,
\ee
for some
$X,Y \in \cG_s$ built of the sources, fields, parameters, and their derivatives. It follows that
one is inevitably led to compute $Im\, \sigma_{\pm}$ and $Ker\, \sigma_{\pm}$, and, moreover,
the cohomology group $H(\sigma_{\pm}) = Ker\,\sigma_{\pm}/Im\, \sigma_{\pm}$ as the operators
$\sigma_{\pm}$ are nilpotent.

By way of example, let us identify independent equations of motion contained  in the
gauge system  \eqref{skelet}-\eqref{bian}. Consider the equations of motion
\eqref{skelet} parameterized by the sources $Z_{(p+1)}(k)$. \footnote{Note that a differential
form $Z_{(p+1)}(k)$ is a tensor product of two groups of indices:  $(p+1)$ antisymmetric world indices
and $k$ totally symmetric traceless fiber indices. For $p=1$ world indices form a singlet, and, therefore,
the tensor product contains a single $o(1,1)$ irreducible component given by totally symmetric traceless tensor.
For $p=0$ the tensor product contains two components given by formula \eqref{dectrl}.}
Those
$o(1,1)$ irreducible components of the sources $Z_{(p+1)}(k)$ that
belong to $Im\, \sigma_{\pm}$ can be shifted to zero by
appropriate shift  redefinitions of fields in terms $\sigma_{\pm}\pi_{(p)}(k\mp1)$ in \eqref{skelet}.
Representing now the Bianchi identity
as \eqref{sigmaeq} one finds that non-vanishing irreducible components of $Z_{(p+1)}(k)$
not belonging  to $Ker\, \sigma_{\pm}$ are auxiliary. That is to say  these components are
expressed through the derivatives of components belonging to the
cohomology $H^{(p+1)}(\sigma_\pm)  = Ker\,\sigma_{\pm}/Im\,
\sigma_{\pm}\big|_{p+1}$, where the slash denotes restriction to
$(p+1)$-forms. Note that cohomology elements of
$H^{(p+1)}(\sigma_\pm)$ represent independent equations of motion
and these  nonetheless are not arbitrary being restricted by the
residual Bianchi identity.

A number of independent identities
between independent equations of motion is equal  to a number of
independent elements of the next cohomology group
$H^{(p+2)}(\sigma_\pm)$. Note that for $p=1$ the Bianchi identities are identically
vanishing $3$-forms and therefore any $2$-from always belongs to
$Ker\, \sigma_{\pm}$. Consequently, there are no differential
constraints in this case and only field redefinitions associated
with $Im \, \sigma_{\pm}$ are possible. These field redefinitions
allow one to shift all non-zero tensors on the right-hand-side of
the unfolded equations \eqref{skelet} to zero except for the
cohomology elements.

Independent fields and gauge parameters can be  considered similarly.  So, the  independent dynamical fields are particular $o(1,1)$ irreducible  components
of $\pi_{(p)}$ identified with elements of $H^{(p)}(\sigma_\pm)$, while other
irreducible $o(1,1)$ components are either auxiliary fields expressed via
dynamical ones, or Stueckelberg fields that can be shifted to zero
by appropriate gauge transformation. Residual gauge parameters are
given by $o(1,1)$ irreducible components identified with elements of
$H^{(p-1)}(\sigma_\pm)$.

In this way, for a given $p = 0,1,2$ we come to the well-known dynamical  interpretation of different cohomology groups
\cite{Shaynkman:2000ts,Bekaert:2005vh,Skvortsov:2009nv} specified to two spacetime dimensions:
\be
\label{interpretation}
\ba{l}
\text{parameters} \in H^{(p-1)}(\sigma_\pm)
\qquad\quad
\text{fields} \in H^{(p)}(\sigma_\pm)
\\
\\
\text{equations} \in H^{(p+1)}(\sigma_\pm)
\qquad\quad\;
\text{identities} \in H^{(p+2)}(\sigma_\pm)
\ea
\ee
All higher cohomology groups are empty,
$H^{(p)}(\sigma_{\pm}) = \varnothing$ for $p\geq 3$, because in $d=2$ dimensions differential $p$-forms with $p\geq 3$ vanish identically. As a corollary, there are no reducible
gauge parameters and identities for identities.

It is important to note that the above  interpretation
of the cohomology elements \eqref{interpretation} gives rise to different forms of one dynamical system
reduced via either  $\sigma_+$ or $\sigma_-$ operators. Generally, this happens because
the respective cohomology groups are non-isomorphic (see below).


\begin{cohomology}
\label{coh_prop_traceless}
\textit{The cohomology groups of operators $\sigma_\pm$ in  $\cG_s$ are given by}
\be
\label{cohprop}
H^{(p)}(\sigma_-) = \left\{
\ba{l}
\dps
p=0:\; T
\\
\\
\dps
p=1, s=1: \; T^{a_1}
\\
\dps
p=1, s>1 :\; T\;, T^{a_1 ... a_{s}}
\\
\\
\dps
p=2: \; T^{a_1 ... a_{s-1}}
\ea
\right.
\;\;\;\;\;\;
H^{(p)}(\sigma_+) = \left\{
\ba{l}
\dps
p=0:\; T^{a_1 ... a_{s-1}}
\\
\\
\dps
p=1, s=1: \; T^{a_1}
\\
\dps
p=1, s>1 :\; T\;, T^{a_1 ... a_{s}}
\\
\\
\dps
p=2: \; T
\ea
\right.
\ee
\textit{where $T^{a_1 ... a_{m}}$ are totally symmetric and traceless $o(1,1)$ tensors.}
\end{cohomology}

\noindent The proof is straightforward and  relegated to Appendix \bref{sec:appendixA}.
\footnote{Our results on $H(\sigma_+)$ cohomology (see a comment in footnote
\bref{f4})
can be obtained from $d$-dimensional consideration of \cite{Skvortsov:2009nv} by taking $d=2$. However,
the case of $d=2$  is strongly degenerate so that making a direct  substitution of $d=2$ should not
be taken for granted. Also, $ H(\sigma_{-})$ has not been discussed  before. In particular,
an explicit computation of the cohomology  has technical features specific to two dimensions that are crucial  when
analyzing the reduced unfolded equations.}
A few comments are in order.

\begin{itemize}
\item  The
cohomology groups establish a cross-duality relation:
\be
\label{cohduality}
H^{(p+2)}(\sigma_\pm) \approx H^{(p)}(\sigma_\mp)\;, \qquad  p=0,1,2; \, (p+2)\; \text{mod 2}\;.
\ee
It underlines dual interpretations of the BF higher spin theory that we develop in the following sections.

\item Elements of group
$H^{(1)}(\sigma_\pm)$ are not double traceless (Fronsdal) spin-$s$
tenors for $s>2$.

\item Scalar elements of $H^{(1)}(\sigma_\pm)$
are two different scalar components of grade $k=1$ element of $\cG_s$,
while tensor components are given by  the same maximally symmetric traceless
component of maximal grade $k=s-1$ element of $\cG_s$ (see Appendix \bref{sec:appendixA} for
more details).

\item Each of the second cohomology groups
$H^{(2)}(\sigma_\pm)$ contains a single non-vanishing element. It
is worth noting that in $d \geq 4$ dimensions  $H^{(2)}(\sigma_-)$
contains two non-vanishing elements called the Einstein cohomology
elements and the Weyl cohomology elements \cite{Vasiliev:2001wa}.
\footnote{See footnote \bref{f4}. In higher spacetime dimensions one  considers
the $\sigma_-$ cohomology only because its elements are interpreted as fields, parameters, and equations
of the Fronsdal theory of massless fields. A dynamical interpretation of the higher spacetime
dimensional  $\sigma_+$ cohomology has not been elaborated yet.} These cohomology elements are
given by differential gauge-invariant combinations of $d$-dimensional Fronsdal
fields and have an elegant interpretation. Indeed, in order to
obtain Fronsdal equations of motion one equates the Einstein
cohomology element to zero, while the Weyl cohomology element
remains arbitrary modulo the Bianchi identities. It follows that the Weyl elements  parameterize on-shell nontrivial gauge
invariant combinations of dynamical fields, \textit{i.e.,} the
physical degrees of freedom. In the  $d=2$ case
$H^{(2)}(\sigma_\pm)$ is spanned by a single element.
\footnote{Along with the second item above this may imply that Fronsdal action
in two dimensions at $s>1$ is a total derivative. \textit{E.g.},
in the $s=2$ case,  the  Einstein tensor does vanish  identically.
On the other hand, the $2d$ Maxwell
action is not a total derivative: the respective
variational equation is $\d_m F=0$, where $F$ stands for
dualized Maxwell tensor. Nonetheless, the theory is topological
because the general solution reads $F=const$ allowing for linear
potentials only.} Equating this
element to zero  inevitably makes the theory topological. We refer elements of $H^{(2)}(\sigma_{\pm})$ to as
the Weyl tensors/scalars.

\end{itemize}

\section{Off-shell unfolded equations for one-form  fields}
\label{sec:offshell}

\paragraph{Component form of fields.} Lorentz components of $0$-form gauge parameters \eqref{transOmega},
1-form gauge fields \eqref{gaugefield}, and 2-form  field strengths \eqref{gaugefield}
will be  denoted as
\be
\xi^{a_1 ... a_k}\;,
\qquad
\omega_m^{a_1... a_k}\;,
\qquad
R_{mn}^{a_1... a_k}\;,
\qquad\;\;
k=0,..., s-1\;;
\ee
all of them satisfy the irreducibility conditions \eqref{symprop}.

Using general formulas \eqref{skelet}, along with \eqref{sigmas} and \eqref{gammakoef}, \eqref{sigmakoef},
we find that the  component form of the field strength  is given by \cite{Alkalaev:2013fsa}
\be
\label{curva}
\ba{l}
\dps
R_{mn}^{a_1 ... a_k}(\omega) = \nabla^{}_{[m} \, \omega_{n]}^{a_1 ... a_k}
-\Lambda \frac{(s-k-1)(s+k)}{2(k+1)} h^{}_{[m}{}_{,\,c} \,\omega_{n]}^{ca_1 ... a_k} +
\\
\\
\dps
\hspace{6cm}+\Big[h_{[m}^{(a_1}\, \, \omega_{n]}^{a_2 ... a_k)}  - \frac{1}{k-1}\, \eta_{}^{(a_1a_2} h^{}_{[m,\,c}\, \,\omega_{n]}^{c a_3 ... a_k)}\Big]
\;.
\ea
\ee

Analogously, the  component form of the gauge symmetry transformations \eqref{transOmega} is given by
\be
\label{omeg}
\ba{l}
\dps
\delta\omega_m^{a_1 ... a_k} = \nabla_m \xi^{a_1 ... a_k} -\Lambda \frac{(s-k-1)(s+k)}{2(k+1)} h_m{}_{,c} \,\xi^{ca_1 ... a_k}+
\\
\\
\dps
\hspace{7cm}+ \Big[h^{(a_1}_m \, \xi^{a_2 ... a_k)}  - \frac{1}{k-1}\, \eta^{(a_1a_2} h_m{}_{,c}\, \xi^{ca_3 ... a_k)}\Big]\;.
\ea
\ee

\paragraph{Off-shell equations of motion.} Consider now  the unfolded equations in the one-form sector
\eqref{1eq} written  in the Lorentz component form as follows
\be
\label{main}
R^{a_1 ... a_k} =  \cV_2 C_{(s)}^{a_1... a_{k}} \;,
\qquad k = 0, ..., s-1\;,
\ee
where $o(1,1)$ totally symmetric and traceless tensors $C_{(s)}^{a_1... a_{k}}$ are the Lorentz  components
of the $2$-form  source $C_{(2)}^{A_1 ... A_{s-1}}$ parameterizing  the right-hand-side of \eqref{1eq}.
The expression  $\cV_2 = \epsilon^{cd} h_c \wedge h_d$ is the volume  2-form (dual to $0$-form)
built of $\ads$ background frame fields.

In the case  $s=1$, the cohomology groups are isomorphic,
$H^{(p)}(\sigma_+) \approx H^{(p)}(\sigma_-)$ for $\forall\, p$.
Therefore, the only equation of motion in \eqref{main} says  that the Maxwell
tensor admits a dual representation, \textit{i.e.}, $R_{mn} \equiv
F_{mn} = \epsilon_{m n} C_{(1)}$. Whence it follows that there are no
restrictions imposed on $F_{mn}$, and the theory is off-shell.
By some means, going on-shell constrains $C_{(1)}$. For instance, by taking  $C_{(1)}=0$ one obtains the BF topological
theory; other possible constraints are discussed in Section
\bref{sec:weyl}. In what follows we always assume $s\geq 2$.

For $s \geq 2$ and $p \neq 1$ the cohomology groups $H^{(p)}(\sigma_\pm)$  are
not isomorphic. This implies that the cohomological reduction of the equation system
\eqref{main} could be done in two different ways giving rise to two different but dynamically
equivalent theories.

Following the general discussion of Section \bref{sec:cohanal}, the unfolded  equations  \eqref{main} can be
represented in two forms depending on particular  operator $\sigma_{\pm}$:
\be
\label{main+}
R^{a_1 ... a_k} = \delta_{k,0} \, \big(C_{(s)}+ \nabla_{b_1} C^{b_1}_{(s)}+\, \cdots\, + \nabla_{b_1} \cdots \nabla_{b_{s-1}}
C_{(s)}^{b_1... b_{s-1}} \big)\cV_2 \;,
\ee
within  the $\sigma_+$ cohomological reduction, and,
\be
\label{main-}
\;\;\; R^{a_1 ... a_k} = \delta_{k,s-1} \big(C_{(s)}^{a_1... a_{s-1}}+ \nabla^{(a_1}C_{(s)}^{a_2 ... a_{s-1})}+ \,\cdots\,
+ \nabla^{(a_1} \cdots \nabla^{a_{s-1})}C_{(s)} + \, \cdots \big)\cV_2\;,
\ee
within the $\sigma_-$ cohomological reduction. In \eqref{main-}
the ellipsis refers to proper symmetrizations of derivatives and trace terms. The proof is analogous to
that of the theorem of Section \bref{sec:cohanal}. The representations \eqref{main+} and \eqref{main-}
are convenient in practice because all field redefinitions have been done
that remove all right-hand-side tensors
$\notin  H^{(2)}(\sigma_\pm)$.
In both cases, we see that field redefinitions produce derivative
transformations setting all the source components to zero except for those
corresponding to the cohomology elements.

The existence of two  operators $\sigma_{\pm}$ used for the
corresponding cohomological reductions  implies two  dual descriptions
of the same system \eqref{main}.
\footnote{Recall that in the flat space  limit $\Lambda = 0$ the
operator $\sigma_-$ disappears (see formula \eqref{curva}) so that
the duality phenomena described below are peculiar to $(A)dS_2$
space only. The cohomological analysis based on the remaining
operator $\sigma_+$ remains valid  in Minkowski space as well.} We
show that the $\sigma_+$ cohomological reduction yields  the
massive scalar Klein-Gordon equation on the hyperboloid with
non-vanishing right-hand-side given by scalar Weyl tensor. The
$\sigma_-$ cohomological reduction yields the  current conservation
condition with non-vanishing right-hand-side given by the higher rank
Weyl tensor. In both cases, we impose partial gauge conditions
setting a part of dynamical  fields to zero.

Recall that the Bianchi identity \eqref{bian} for the equation
system \eqref{main} is trivial thereby implying that the cohomology
elements are arbitrary.  Imposing  algebraic and/or differential
constraint on Weyl scalars/tensors is discussed in Section
\bref{sec:weyl}. For instance, equating all the cohomology elements to zero  one obtains the BF
higher spin theory with the action \eqref{actBFlin}.

\subsection{Explicit $\sigma_+$ - reduction: one-form sector  }
\label{sec:sigmaplus}

For convenience, we use the
representation \eqref{main+} with   $C_{(s)}^{b_1... b_{k}} = 0$, where $k =  1,2,..., s-1$. It follows that the unfolded equations take the form
\be
\label{constr2}
R^{} = \cV_2 C^{(s)}\;,  \quad R^{a_1 ...\, a_k} = 0\;,
\qquad k=1,...,s-1\;,
\ee
where Weyl scalar $C^{(s)} \in H^{(2)}(\sigma_+)$ is arbitrary function of spacetime variables,
and the field strengths $R^{a_1 ... \, a_k}(\omega)$ are given by \eqref{curva}.

The cohomological approach says that the field space $\cG_s$  in the sector of $1$-form fields $\omega_m^{a_1 ... \, a_k}$
decomposes into Stueckelberg fields, auxiliary fields, and dynamical fields given by the cohomology $H^{(1)}(\sigma_+)$.
The above three types of  fields appear as particular irreducible Lorentz components of $\omega_m^{a_1 ... \, a_k}$,
cf. \eqref{dectrl}.

In the case $s>1$, the vanishing higher rank field strengths  at $k\neq 0$ are constraints allowing
to express auxiliary fields via derivatives of independent dynamical fields given  by a scalar and a rank-$s$
traceless tensor $\varphi, \varphi^{a_1 ... a_s} \in H^{(1)}(\sigma_+)$ \eqref{cohprop}. Other Lorentz components of $\omega_m^{a_1 ... \, a_k}$
are Stueckelberg ones shifted to zero by algebraic parts of the gauge transformations \eqref{omeg}.

The minimal grade degree equation $\epsilon^{mn}R_{mn} = C^{(s)}$ is the only off-shell
equation of motion for dynamical  fields.
Gauge fixing all Stueckelberg fields to zero and expressing all auxiliary fields via the
dynamical fields, one shows that the minimal grade equation is reduced to the following
order-$s$ differential equation
\be
\label{scal}
\kappa_s\,\big( \epsilon_{a_1b}\nabla^{b}\nabla_{a_2} ... \nabla_{a_s} \varphi^{a_1 a_2 ... a_s}\big)
+  \rho_s\big(\Box_{_{\ads}}  + m_s^2\big)\varphi = C^{(s)}\;,
\ee
with
\be
\label{mass}
m_s^2 = 2\rho_0 \equiv -s(s-1)\Lambda\;,\quad s\geq 2\;,
\ee
where $\Box_{_{\ads}} = \nabla^a \nabla_a$  is the wave operator on the $\ads$ background, and
coefficient $\rho_0$ is given by \eqref{sigmakoef}. Non-vanishing spin-dependent coefficients $\kappa_s, \rho_s$ are fixed by gauge symmetry
transformations
\be
\label{gauge1}
\delta \varphi^{a_1 ... a_s} = \nabla^{(a_1} \xi^{a_2 ... a_{s})} -\frac{1}{s-1} \;\eta^{(a_1a_2} \nabla_c\, \xi^{a_3 ... a_{s-1})c}\;,
\ee

\be
\label{gauge2}
\delta \varphi = \epsilon_{ba_1}\nabla^b \nabla_{a_2} \cdots \nabla_{a_{s-1}}\xi^{a_1 a_2 \cdots\, a_{s-1}}\;,
\ee
with an independent gauge parameter $\xi^{a_1 ... a_{s-1}}\in H^{(0)}(\sigma_+)$, see \eqref{cohprop}.
Lower grade degree $k=0,1,..., s-2$ gauge parameters  $\xi^{a_1 ... a_k}$ are Stueckelberg ones used to
shift some Lorentz components in $\omega_m^{a_1 ... \, a_k}$ to zero.

The dynamical equation \eqref{scal} can be simplified. To this end, a field $\varphi^{a_1 ... a_s}$
is completely gauged away by imposing the higher-spin gauge
\be
\label{hsgc}
\varphi^{a_1 ...\, a_s} = 0\;.
\ee
Indeed, a traceless rank-$k$ tensor in $d=2$ dimensions  has two independent components for any $k\geq 2$.
It follows that  a number of independent components of a
rank-$(s-1)$ gauge parameter equals a number of equations in \eqref{hsgc}.
The higher spin gauge can be viewed as an extension of the standard conformal gauge in $2d$
gravity which  makes the metric proportional to Minkowski tensor.
Then, the only dynamical field is given by a scalar component of the cohomology group,  $\varphi \in H^{(1)}(\sigma_+)$.

Imposing the higher spin gauge \eqref{hsgc} and solving the constraints in  \eqref{constr2} one finds that
the leftover equation reduces to the massive scalar equation with particular value
of the mass-like term \cite{Alkalaev:2013fsa}
\be
\label{scaleq}
\Box_{_{\ads}}\varphi  -s(s-1)\Lambda\varphi  = C^{(s)}\;,
\ee
where we redefined the right-hand-side as $\rho_s^{-1} C^{(s)} \rightarrow  C^{(s)} $.

The massive  scalar equation  \eqref{scaleq} is invariant with
respect to residual gauge transformations \eqref{gauge2}
provided that the gauge parameter $\xi^{a_1 ... a_{s-1}} \in H^{(0)}(\sigma_+)$, satisfies   the generalized
 Killing equation on the hyperboloid,
\be
\label{CKE}
\nabla^{(a_1} \xi^{a_2 ... a_{s})} -\frac{1}{s-1} \;\eta^{(a_1a_2} \nabla_c\, \xi^{a_3 ... a_{s-1})c} = 0\;,
\ee
The above constraint  is clearly explained
as the stability transformation of the higher spin gauge condition \eqref{hsgc}  for transformations \eqref{gauge1}.

A few comments are in order.

\begin{itemize}

\item In the spin-$2$ case
the above equation reproduces the gauge-fixed linearized equation of motion of the Jackiw-Teitelboim model
in the one-form sector \cite{Fukuyama:1985gg,Jackiw:1992bw, Alkalaev:2013fsa}.
We see that the higher spin extension is described by the scalar field as well, but with a different spin-dependent
mass term \eqref{mass} and higher derivative leftover gauge symmetry \eqref{CKE}.

\item Mass  $m_s^2$ \eqref{mass} differs from the conformal
value of mass $m_{conf}^2 = -\Lambda\, d(d-2)/4 = 0$ in $d=2$ dimensions.

\item Mass $m_s^2$ coincides with the value of the Casimir operator
of $o(2,1)$ global symmetry algebra of $\ads$ space realized on tensor fields .

\item Since the theory propagates no local degrees of freedom,  the scalar  field equation
\eqref{scaleq} at $C^{(s)} \neq 0$  becomes  a constraint equation  for auxiliary field $\varphi$ that can be solved by defining
the respective Green's function: $\varphi(x) = \dps(\Box_{\ads} + m_s^2)^{-1} C^{(s)}(x)$.

\end{itemize}

\subsection{Explicit $\sigma_-$ - reduction: one-form sector    }
\label{sec:sigmaminus}

Using the representation \eqref{main-} with  $C_{(s)}^{b_1... b_{k}} = 0$, $k =  0,1,..., s-2$, one
arrives at the following unfolded equations
\be
\label{constr1}
R^{a_1 ... a_{s-1}} = \cV_2 C^{a_1 ... a_{s-1}}\;, \quad R^{a_1 ... a_k} = 0 \;,\qquad  k  = 0,..., s-2\;,
\ee
where Weyl tensor  $C^{a_1 ... a_{s-1}}_{(s)} \in H^{(2)}(\sigma_-)$ is arbitrary function of spacetime variables,
and the field strengths $R^{a_1 ... \, a_k}(\omega)$ are given by \eqref{curva}.

In the case $s>1$, the vanishing higher rank field strengths  at $ k  = 0, ..., s-2$  are constraints allowing
to express auxiliary fields via derivatives of independent dynamical fields given  by a scalar and a rank-$s$
traceless tensor $\phi, \phi^{a_1 ... a_s} \in H^{(1)}(\sigma_+)$ \eqref{cohprop}. Other Lorentz components of $\omega_m^{a_1 ... \, a_k}$
are Stueckelberg ones shifted to zero by algebraic parts of the gauge transformations \eqref{omeg}.

Solving  the constraints  \eqref{constr1} yields
the following expression
\be
\label{solvco}
\omega_{m|a_1...a_{s-1}}  = \phi_{ma_1 ... a_{s-1}} + \tau_s \big( \, \eta_{ma_1} \nabla_{a_2} ... \nabla_{a_{s-1}} \phi + ...\,\big)\;,
\ee
where $\tau_s$ is some non-vanishing spin-dependent coefficient,
the parenthesis contain terms that depend on field $\phi$ only, while  the ellipsis refers to
appropriate symmetrizations of derivatives and trace terms.
Independent gauge  transformations are  given by
\be
\label{residualtr1}
\delta \phi =  \big(\Box_{_{\ads}}  + m_s^2 \big)\xi\;,
\ee

\be
\label{residualtr2}
\delta \phi_{a_1 ... a_{s}} = \frac{1}{\Lambda}\,\nabla_{a_1} \cdots  \nabla_{a_s} \xi + ...\;,
\ee
where the ellipsis refers to proper symmetrizations and trace terms, while a scalar gauge parameter
$\xi \in H^{(0)}(\sigma_-)$ \eqref{cohprop}.  The mass coefficient $m^2_s$ is given by \eqref{mass}.

The maximal  grade degree equation $R^{a_1 ... a_{s-1}} = \cV_2 C^{a_1 ... a_{s-1}}$ is the only off-shell
equation of motion for dynamical  fields.
Gauge fixing all Stueckelberg fields to zero and expressing all auxiliary fields via the
dynamical fields using  \eqref{solvco}, one shows that the maximal grade equation is reduced to the following
order-$(s-1)$ differential equation
\be
\label{tenso}
\nabla^m \phi_{ma_1 ... a_{s-1}} - \tau_s \,\nabla _{a_1} ... \nabla_{a_{s-1}} \phi + \ldots = C^{(s)}_{a_1 ... a_{s-1}}\;,
\ee
where the ellipsis refers to proper symmetrizations and trace terms.

Higher order equation \eqref{tenso} can be simplified by imposing a gauge condition.
Indeed, using the scalar field  transformations \eqref{residualtr1} one introduces  the
scalar gauge condition along with the residual gauge parameter equation
\be
\label{hsgc-}
\phi= 0\;,
\qquad
\Box_{_{\ads}} \xi -m_s^2 \xi =0\;,
\ee
which are dual cousins of higher spin gauge condition \eqref{hsgc}  and generalized Killing equations \eqref{CKE}.
It follows that dynamical equation \eqref{tenso} takes the form
\be
\label{inteq}
\nabla^n \phi_{n a_1 ... a_{s-1}} = C^{(s)}_{a_1 ... a_{s-1}}\;.
\ee

For equation \eqref{inteq} with the vanishing right-hand-side $C^{(s)}_{a_1 ... a_{s-1}} =0$ one identifies
$\phi_{a_1 ... a_{s}}$ with spin-$s$ conserved current on the hyperboloid.
\footnote{For particular models, switching on non-vanishing tensors on the right-hand-side
may be visualized as a sort of covariantization characteristic to non-Abelian interaction theories,
which therefore is not conservation violation but rather a map $\nabla_m \rightarrow D_m$, where $D_m$
is some new field-dependent covariant derivative.}
Higher order derivative transformations \eqref{residualtr2} with the scalar gauge parameter satisfying
\eqref{hsgc-}
are treated now as "improvement" transformations for conserved currents. Indeed, "improvements"
are higher order derivative transformations with an antisymmetric
tensor parameter which in $d=2$ dimensions is dualized  to a scalar via the Levi-Civita tensor.

Our analysis of the $\sigma_-$ cohomological   reduction applied to the unfolded equations in the one-form sector
yields  the following interpretation of the cohomology groups $H^{(p)}(\sigma_-)$,
which conforms the general scheme \eqref{interpretation}.  Namely, elements
$C^{(s)}_{a_1 ... a_{s-1}} \in H^{(2)}(\sigma_-)$ are conservation conditions.
Element $\phi \in H^{(1)}(\sigma_-)$ can be chosen a pure gauge, so that another
cohomology element $\phi_{a_1 ... a_{s}} \in H^{(1)}(\sigma_-)$ can be identified with a conserved current.
Element $\xi \in H^{(0)}(\sigma_-)$ plays the role of an "improvement" transformation parameter.

\subsection{Off-shell field spaces}
\label{sec:weyl}

In the framework of  the unfolded formulation one may  introduce the so-called  Weyl module as a linear space which elements
parameterize all possible gauge-invariant differential combinations of dynamical fields $\in H^{(1)}(\sigma_{\pm})$ that remain arbitrary on-shell. In   $d\geq 4 $
dimensions, the Weyl module is derived by solving the Bianchi identities: one "unfolds" the original higher
spin Weyl tensor, \textit{i.e.}  introduces new variables (infinite of them) that parameterize independent
combinations of derivatives of the Weyl tensor \cite{Bekaert:2005vh}.

In $d=2$ dimensions the Bianchi
identities in the one-form sector trivialize due  to $H^{(3)}(\sigma_{\pm})= \varnothing$,
see \eqref{bian} and \eqref{interpretation}. Whence, the Weyl tensor $\in H^{(2)}(\sigma_{\pm})$ remains completely
arbitrary function of spacetime variables. However, it does not yield local
degrees of freedom in the theory. Indeed, recall that contrary to the higher-dimensional case,
the cohomology $H^{(2)}(\sigma_{\pm})$ contains the only element, cf. \eqref{cohprop}. In other words,
the Einstein cohomology (higher spin equations of motion) and the Weyl cohomology (higher spin Weyl tensors) coincide in two dimensions. It follows that
keeping the Weyl element arbitrary implies the theory is off-shell. On the other hand, choosing
the Weyl element to be a particular function  can be treated as "going on-shell". \textit{E.g.},
setting all Weyl tensors to zero results in the  zero-curvature equations of motion   \eqref{linBFeom}. There are
various ways of how to put our topological system on-shell. We discuss some of them
in  Section \bref{sec:onshell}.

\subsubsection{Unfolding Weyl tensors}

Despite the lack  of $2d$ Bianchi identities,  one can  still associate to Weyl tensors
infinite sets of components which comprise their all possible
derivative combinations.
Namely, by off-shell  field space  for the Weyl scalar $C^{(s)} \in H^{(2)}(\sigma_{+})$
we call the following set of components
\be
\label{scalaroffshell0}
\mathcal{W}_0= \big\{W^{(s)}_{b_1...\, b_k}\;,\quad k=0,1,2,... \big\}\;,
\ee
where elements are totally symmetric and traceful, $\eta^{mn}\, W^{(s)}_{mn\,b_1...\, b_{k-2}} \neq 0$
for $k = 2,3,...$, so that one identifies an index-free component with the original Weyl scalar,  $W^{(s)} \equiv C^{(s)}$.
Elements of $\cW_0$ are equated  with all possible  derivatives of
original scalar $C^{(s)}$, \textit{i.e.},
\be
\label{offshelconstr1}
W^{(s)}_{b_1...\, b_k}  - \cP_{b_1...\, b_k} C^{(s)} = 0\;, \qquad \cP_{b_1...\, b_k} = \nabla_{b_1} \cdots \nabla_{b_k} + \cdots\;,
\ee
where the ellipses in \eqref{offshelconstr1}
refers to proper symmetrizations and all possible trace terms. For a given $k$ the projector $\cP_{b_1...\, b_k}$ contains a finitely many
arbitrary coefficients not fixed by the above definition of $\cW_0$.
Note that in $d=2$ dimensions only symmetric combinations of covariant derivatives are possible
because any non-symmetric $\nabla^{a_1} ... \nabla^{a_k} C$ can be reduced to a collection of
symmetrized combinations by using  the Levi-Civita tensor and commutator $[\nabla, \nabla] \sim \Lambda$.

Quite analogously, by off-shell field space  for  the Weyl tensor $C_{a_1 ... a_{s-1}}^{(s)} \in H^{(2)}(\sigma_{-})$
we call the following set of components
\be
\label{scalaroffshells}
\mathcal{W}_{s-1}= \big\{W^{(s)}_{a_1 ... a_{s-1} | b_1...\, b_k}\;,\quad k=0,1,2,... \big\}\;,
\ee
where  elements are totally symmetric in each group of indices, and traceless with respect to the first group of  indices,
$\eta^{mn}\, W^{(s)}_{mn a_1 ... a_{s-3} |b_1 ...\, b_k} = 0$, and traceful with respect to
the second group of  indices,
$\eta^{mn}\, W^{(s)}_{a_1 ... a_{s-1} |b_1 ...\, b_{k-2}mn} \neq 0$. The $k=0$ element is identified
with the original Weyl tensor, $W^{(s)}_{a_1 ... a_{s-1}} \equiv C^{(s)}_{a_1 ... a_{s-1}}$.
Elements of $\cW_{s-1}$ are equated  with all possible  derivatives of
original tensor  $C^{(s)}_{a_1 ... a_{s-1}}$, \textit{i.e.},
\be
W^{(s)}_{a_1 ... a_s | b_1...\, b_k}  - \cP_{b_1...\, b_k} C^{(s)}_{a_1 ... a_{s-1}} = 0\;,
\qquad \cP_{b_1...\, b_k} = \nabla_{b_1} \cdots \nabla_{b_k} + \cdots\;.
\ee

Generally, off-shell field space elements  are not related to each other.  A natural option suggested  in
\cite{Shaynkman:2000ts} is to consider particular constraints for elements of the off-shell field space
relating  components with different values of
$k$ as
\be
\label{ShV}
W^{(s)}_{b_1 ... b_{k+1}} \sim \nabla_{b_1} W^{(s)}_{b_2... b_{k+1}}\;,
\ee
while element $W^{(s)}$ remains arbitrary. It follows
that the form of relations \eqref{offshelconstr1} is not changed, while arbitrary coefficients in
projectors $\cP_{b_1...\, b_k}$ are uniquely fixed modulo a single free coefficient to be  identified
with the mass parameter. We refer the off-shell field space $\cW_0$ supplemented with constraints \eqref{ShV} to as the off-shell Weyl
module ${\widetilde\cW}_0$. The same consideration can be applied to off-shell module $\cW_{s-1}$.

\subsubsection{Going on-shell}
\label{sec:onshell}

Recall now that dynamical fields propagated by the unfolded
equations \eqref{main} are considered as auxiliary, see our
comments in the end of  Section \bref{sec:sigmaplus}. Indeed, these are
completely expressed via the Weyl tensors which parameterize the
right-hand-sides of the dynamical equations. Such a phenomenon is
characteristic of topological field theories coupled to external
dynamical systems with or without local degrees of freedom (see a
recent discussion in \cite{Kapustin:2014gua}). In particular, this
is the way one couples matter fields to $3d$ topological
Chern-Simons theory. In this case, Chern-Simons strength tensor
turns out to be proportional to a matter current so that
respective gauge fields are auxiliary carrying no physical degrees
of freedom. However, added topological modes may have a profound
impact on dynamics of the matter system, giving rise, for
instance, to anyonic statistics.

In our case, the problem of coupling  a field theory with an
(in)finite number of degrees of freedom to the  topological
unfolded theory given by equations \eqref{main} reduces to the
equivalent problem of specifying  Weyl tensors via  imposing
appropriate constraints on elements of the off-shell field spaces.
Note that  choosing particular  Weyl tensors actually
puts  the topological system  \eqref{main} on-shell. Other way round,
going on-shell in the topological theory \eqref{main} is nicely interpreted as
coupling to external field theory.

By way of example, specify the off-shell field space $\cW_0$ to the off-shell Weyl module ${\widetilde\cW}_0$ given by
\eqref{ShV}, and
impose the tracelessness condition
\be
\label{constraint1}
\eta^{mn}\, W^{(s)}_{mn\, b_3...\, b_k}  = 0 \;.
\ee
The above constraint yields the massive Klein-Gordon equation of motion on $\ads$ spacetime imposed
on the Weyl scalar $C^{(s)}$ \cite{Vasiliev:1995sv,Shaynkman:2000ts}. It follows that
an external field theory is identified here  as the scalar field  theory coupled to (linearized) topological spin-$s$ BF
theory. The dynamical field $\varphi$ in
equation \eqref{scaleq} is auxiliary and expresses now in terms of the Klein-Gordon field $C^{(s)}$.

As another possible option let us  mention a truncation of the off-shell Weyl ${\widetilde\cW}_0$ by imposing the following constraint
\be
\label{constraint2}
W^{(s)}_{b_1 ... b_k} = 0\; \quad  \text{for}\quad  \; k = m, m+1, ..., \infty\;,
\ee
at some fixed $m$. The above truncation is most easily analyzed in the spin $s=1$ case.
Here, there are two standard choices of $m=1$ and $m=0$. Truncating $\cW_0$  by
imposing $W^{(1)}_b =0$ is equivalent to   constraint  $\nabla_b
F = 0$ which is the dualized Maxwell  equation. Recall here that  dualized Maxwell tensor $F_{mn} =
\epsilon_{mn}F$ is identified with scalar $C^{(1)}$ and two off-shell field spaces
considered above coincide, being actually a single space $\cW_0$. Also, one may
truncate all elements of $\cW_0$ by imposing constraint $W^{(1)} \equiv F =
0$ that appears as the equation of motion  in
the Abelian BF theory.


%

Another example of a theory with no local degrees of freedom identified with an external field theory is given by
equations   \eqref{2eq}-\eqref{1eq} with the right-hand-sides given by  \eqref{augment}. In this case, the right-hand-side of
unfolded equation \eqref{main} is parameterized by $0$-form field subjected to another unfolded equation
which describes no local degrees of freedom as well (see the next section).

\section{Off-shell unfolded equations for zero-form fields}
\label{sec:0forms}

Consider  now the unfolded equations in the zero-form sector \eqref{skelet}.
By analogy with \eqref{dectraceless}  $o(2,1)$ covariant $0$-form fields can be decomposed
into Lorentz algebra $o(1,1) \subset o(2,1)$  components as
\be
\label{dectracelessPHI}
\Phi^{A_1 ... \, A_{s-1}} = \bigoplus_{k=0}^{s-1}\;\phi{}^{a_1 ... \, a_k}\;,
\ee
where Lorentz  components satisfy irreducibility conditions \eqref{symprop}.
Using general formulas \eqref{skelet}, along with \eqref{sigmas} and \eqref{gammakoef}, \eqref{sigmakoef},
one finds that Lorentz  component form of equations \eqref{skelet}  reads as
\be
\label{main3}
D^{a_1 ... a_k |m} = B^{a_1 ... a_k|m}\;,
\qquad
k = 0,..., s-1\;,
\ee
where
\be
\label{Dcurva}
\ba{l}
\dps
D^{a_1 ... a_k}_m = \nabla_m \phi^{a_1 ... a_k} -\Lambda \frac{(s-k-1)(s+k)}{2(k+1)} h_m{}_{,c} \,\phi^{ca_1 ... a_k}+
\\
\\
\dps
\hspace{6cm}+ \Big[h_m^{(a_1} \, \phi^{a_2 ... a_k)}  - \frac{1}{k-1}\, \eta^{(a_1a_2} h_m{}_{,c}\, \phi^{ca_3 ... a_k)}\Big]\;,
\ea
\ee
where $D^{a_1 ... a_k |m} = h^{m,n}D_n^{a_1 ... a_k}$ and the slash
says  that two groups of fiber indices are not related by permutations, tensors $B^{a_1 ... a_k|m}$
are $o(1,1)$ components of  differential $1$-form $B_{(1)}^{A_1 ... A_{s-1}}$ \eqref{2eq}.

The $0$-form fields have no associated gauge symmetry \eqref{transPhi}. However,
the equations of motion \eqref{main3}  satisfy the Bianchi identities taking the following component
form, cf. \eqref{bian},
\be
\label{idenD}
\ba{l}
\dps
\nabla^{}_{[m} \, D_{n]}^{a_1 ... a_k}
-\Lambda \frac{(s-k-1)(s+k)}{2(k+1)} h^{}_{[m}{}_{,\,c} \,D_{n]}^{ca_1 ... a_k} +
\\
\\
\dps
\hspace{5cm}+\Big[h_{[m}^{(a_1}\, \, D_{n]}^{a_2 ... a_k)}  - \frac{1}{k-1}\, \eta_{}^{(a_1a_2} h^{}_{[m,\,c}\, \,D_{n]}^{c a_3 ... a_k)}\Big]
 \equiv 0\;.
\ea
\ee

According to the general consideration of Section
\bref{sec:cohanal}, the system \eqref{main3} can be algebraically
reduced using one or another type of nilpotent operators
$\sigma_\pm$. In both cases, the cohomological theorem \eqref{cohprop}
guarantees that the true dynamical fields in the system \label{main2}
are  either $\phi \in H^{(0)}(\sigma_-) $, or
 $\phi^{a_1 ... a_{s-1}} \in H^{(0)}(\sigma_+)$.
Cohomology elements $B^{\pm(s)}, \,B^{\pm(s)}_{a_1... a_{s}} \in H^{(1)}(\sigma_\pm)$ represent
independent equations of motion. A number of independent
identities between equations of motion corresponds to a number of independent elements of the second
cohomology group, \textit{i.e.,}  $I^{(s)}_{a_1 ... a_{s-1}} \in H^{(2)}(\sigma_-)$ and $I^{(s)} \in H^{(2)}(\sigma_+)$.

Note that the right-hand side of the equation system \eqref{main3} cannot
be set to  $\delta_{k,1} \, (\epsilon_{ma_1} B^{+(s)} +
\eta_{ma_1}B^{-(s)})  + \delta_{k,s-1}(B_{a_1... a_{s-1}m}^{+(s)}
+B_{a_1... a_{s-1}m}^{-(s)})$ as in the
case of the unfolded equations in the one-form sector \eqref{main}. Not only the cohomology elements,
but also other components $B^{a_1 ... a_k|m}$ are generally
non-vanishing. While the cohomology represents the independent equations of motion,
the other components are auxiliary, \textit{i.e.}, are expressed through the independent ones by
virtue of the Bianchi identities, see Section  \bref{sec:cohanal}.

It is worth noting that the right-hand-sides of the
independent equations of motion obtained through the cohomological reduction  are parameterized by two independent elements
of $H^{(1)}(\sigma_\pm)$. In this respect, the situation is
different from that in the gauge sector, where the reduced equations of motion are parameterized by a single
Weyl scalar/tensor.  It is
similar to the higher dimensional picture, where the right-hand-sides
of the equations also contain  two independent cohomology elements,
the Einstein part and the Weyl part, see the discussion in the end of Section
\bref{sec:cohanal}. \footnote{It would be instructive  to explicitly build Weyl-like linear spaces
that parameterize  solutions to the Bianchi identities. See our discussion of the off-shell field spaces in the gauge sector
in Section \bref{sec:weyl}.}

\subsection{Explicit $\sigma_+$ - reduction: zero-form sector  }

The $\sigma_+$ cohomological reduction of the unfolded equations \eqref{main3} gives rise to the following independent
equations of motion
\be
\label{pon}
\ba{c}
\dps
\epsilon_{mn_1}\nabla^m \nabla_{n_2} \cdots \nabla_{n_{s-1}}\varphi^{n_1 \cdots\, n_{s-1}}  = B^{+(s)}\;,
\\
\\
\dps
\nabla_{(a_1} \varphi_{a_2 ... a_{s})} -\frac{1}{s-1} \;\eta_{(a_1a_2} \nabla^c\, \varphi_{a_3 ... a_{s-1})c}
= B_{a_1... a_{s}}^{+(s)}\;,
\ea
\ee
where $\varphi_{a_1 \cdots\, a_{s-1}} \in H^{(0)}(\sigma_+)$ and $B^{+(s)}, B_{a_1.... a_{s}}^{+(s)} \in
H^{(1)}(\sigma_+)$, and indices are symmetrized with a unit weight. The tensors on the right-hand-sides of \eqref{pon}
are not arbitrary and are subjected to the Bianchi identities \eqref{idenD}. Following \eqref{interpretation} and \eqref{cohprop},
we find that there is a single identity between independent equations \eqref{pon} corresponding
to a scalar element $I^{(s)}\in H^{(2)}(\sigma_+)$,
\be
\label{idenDmetr}
\kappa_s\,\big( \epsilon^{a_1b}\nabla_{b}\nabla^{a_2} ... \nabla^{a_s} B^{+(s)}_{a_1 a_2 ... a_s}\big)
+  \rho_s\big(\Box_{_{\ads}}  + m_s^2\big)B^{+(s)} = 0\;,
\ee
where $\kappa_s, \rho_s$ are some non-vanishing spin-dependent coefficients, cf. \eqref{scal},
while mass parameter $m^2_s$ is given by \eqref{mass}.

By way of example, consider the spin-$2$ case. Here, the unfolded equations of motion \eqref{main3}
read
\be
\label{2spin2eq}
\nabla_m \varphi - \Lambda  h_m^c \varphi_c  = B_m\;,
\qquad
\nabla_m \varphi^a  + h_m^a \varphi = B_m^a\;,
\ee
where $\Lambda$ is the cosmological constant, and  $B$ and $B^a$ are subjected to
the Bianchi identities \eqref{idenD}
\be
\label{2spin2bian}
\nabla_m B  - \Lambda h_m^c B_c = 0\;,
\qquad
\nabla_m B^a  + h_m^a B = 0\;.
\ee
Using the $\sigma_+$ cohomological reduction and  field redefinitions one finds
from the second equation in \eqref{2spin2eq} that  $\varphi =  -\frac{1}{2}
\nabla^a \varphi_a$. Considering the Bianchi identities
\eqref{2spin2bian} one shows that the first equation in
\eqref{2spin2eq} is a  differential consequence of the  second
equation. The resulting  equations that follow from the
second equation in \eqref{2spin2eq} for the independent field
$\varphi^a \in H^{(0)}(\sigma_+)$ read
\be
\label{B+}
\nabla_a \varphi_b + \nabla_b\varphi_a - \eta_{ab}\nabla^c\varphi_c = B_{ab}^{+}\;,
\qquad \epsilon^{ab}\nabla_a \varphi_b = B_+\;,
\ee
where $B^+, B_{(ab)}^+ \in
H^{(1)}(\sigma_+)$; cf. equations \eqref{pon}.  Note that redefining fields by a dualization
via $\epsilon^{ab}$-tensor  yields
the following system $\nabla_a \varphi_b + \nabla_b\varphi_a =
B_{ab}$, where $B_{ab} = B^+_{ab} + \epsilon_{ab}B^+$,
\footnote{Here we used formula \eqref{decspin2}. The trace component is set to zero by a shift field redefinition
because it belongs to $Im\, \sigma_+$.}. This form is useful when analyzing Killing symmetries
of the gauge dynamical field, see Section \bref{sec:globsym}.
The Bianchi identities \eqref{idenDmetr} take the form
\be
\epsilon^{ac} \nabla_c \nabla^{b} B^+_{a b} + \big(\Box_{_{\ads}}  -2 \Lambda\big)B^{+} = 0\;,
\ee
or, equivalently, $\epsilon_{ab}\big(\nabla^a \nabla^c B^b{}_c  + \Lambda B^{ab}\big)= 0$.  We see that
there is a single identity corresponding to a single element of the second cohomology
$I \in H^{(2)}(\sigma_+)$.

\subsection{Explicit $\sigma_-$ - reduction: zero-form sector }

The $\sigma_-$ cohomological reduction of the unfolded equations \eqref{main3} gives rise to the following independent
equations of motion
\be
\label{vto}
\ba{c}
\dps
\big(\Box_{_{\ads}}   + m^2_s\big)\phi  = B^{-(s)}\;,
\\
\\
\dps
\big(\nabla_{a_1} \cdots \nabla_{a_{s}} \phi + ...  \big)=
B_{a_1... a_{s}}^{-(s)}\;,
\ea
\ee
where $\phi \in H^{(0)}(\sigma_-)$ and $B^{-(s)}, B^{-(s)}_{a_1.... a_{s}} \in
H^{(1)}(\sigma_-)$, coefficient $m_s^2$ is given by \eqref{mass};
the ellipses refers to proper symmetrizations and trace terms.
The right-hand-sides of equations \eqref{vto} are not arbitrary and are subjected to the Bianchi identities
\eqref{idenD}. Following \eqref{interpretation} and \eqref{cohprop},
we find that there is a tensor identity between independent equations \eqref{vto} corresponding
to a tensor element $I_{a_1 ... a_{s-1}}^{(s)}\in H^{(2)}(\sigma_-)$,
\be
\label{Bianchi3}
\nabla^n B^{-(s)}_{na_1 ... a_{s-1}}
-  \tau_s \big( \nabla _{a_1} ... \nabla_{a_{s-1}} B^{-(s)} + \ldots\big) = 0\;,
\ee
where $\tau_s$ is some non-vanishing spin-dependent coefficients, cf. \eqref{scal};
the ellipses refers to proper symmetrizations and trace terms.

By way of example, consider the spin-$2$ case. Here, the equations of motion and the Bianchi identities are
the same as in the previous section, see \eqref{2spin2eq} and \eqref{2spin2bian}.
The cohomological analysis goes along the same lines.
So, using the $\sigma_-$ cohomological  reduction one finds
from the first equation in \eqref{2spin2eq} that $\phi^a =  - \nabla^a \phi$. It follows that
the resulting   equation for
the independent field $\phi \in H^{(0)}(\sigma_-)$ reads $\nabla_a \nabla_b\phi - \eta_{ab}\Lambda\phi = B_{ab}$,
where tensor $B_{ab} = B^-_{ab} + \eta_{ab}B^-$,
while  $B^-, B^-_{ab} \in H^{(1)}(\sigma_-)$. The trace and traceless parts of the above equation
are
\be
\label{vtor}
\Box_{_{\ads}}\phi -2\Lambda \phi = B^-\;,
\qquad
\nabla_a \nabla_b\phi - \half \eta_{ab}\,\Box_{_{\ads}} \phi = B^-_{ab}\;,
\ee
cf. equations \eqref{vto}. Equations \eqref{vtor} reproduce the Jackiw-Teitelboim
linearized equations in the zero-form sector \cite{Fukuyama:1985gg}.
The  Bianchi identities \eqref{Bianchi3}
take the form
\be
\nabla^b B_{ab}^{-} - \nabla_a B^{-} = 0\;,
\ee
or, equivalently, $\epsilon^{bc}\nabla_b B_c{}^a = 0$. We see that
there is an $o(1,1)$ vector identity corresponding to independent elements of the second cohomology
$I^a \in H^{(2)}(\sigma_-)$.

\subsection{Background symmetries}
\label{sec:globsym}

The unfolded equations in the zero-form sector  \eqref{2eq} can be considered from a different perspective.
Provided the right-hand-side is vanishing,  the equations \eqref{2eq} are interpreted as
stability transformations for a particular  $1$-form background gauge field $\Omega_0$. From \eqref{transOmega} it follows that the stability transformation equation
reads
\be
\label{transstab1}
D_0 \xi^{A_1 ... A_{s-1}} = 0\;,
\ee
while its $o(1,1)$ component form read off from \eqref{omeg}  is given by
\be
\label{transstab2}
\ba{l}
\dps
\nabla_m \xi^{a_1 ... a_k} -\Lambda \frac{(s-k-1)(s+k)}{2(k+1)} h_m{}_{,c} \,\xi^{ca_1 ... a_k}+
\\
\\
\dps
\hspace{6cm}+ \Big[h^{(a_1}_m \, \xi^{a_2 ... a_k)}  - \frac{1}{k-1}\, \eta^{(a_1a_2} h_m{}_{,c}\, \xi^{ca_3 ... a_k)}\Big] = 0\;.
\ea
\ee

Taking into account the analysis of the unfolded equations in the zero-form sector,
the system \eqref{transstab2} can be treated in two different ways, using either $\sigma_-$ or $\sigma_+$
cohomological reduction. Whence it follows that  there are two possible interpretations of the stability
transformations.

Using the $\sigma_+$ cohomological reduction one finds out that  \eqref{transstab2} reduces to equations
\eqref{gauge1}-\eqref{gauge2} or \eqref{pon} on tensor parameters $\xi^{a_1 ... a_{s-1}}$ at  $s = 1,2,..., \infty$
subjected to the  Bianchi identity \eqref{idenDmetr}. For a given $s$, the solution to the stability
equations  depends on  a finitely many integration constants interpreted as
constant $o(1,1)$ tensors parameterizing higher spin  global symmetry transformations of the
$\ads$ background spacetime. \footnote{Detailed discussion of global higher spin symmetries in higher  dimensions
and their representations can be found, \textit{e.g.},
in \cite{Vasiliev:1999ba,Konstein:1989ij,Bekaert:2007mi, Boulanger:2013zza,Joung:2014qya}.}
For instance, in the spin-$2$ case stability transformation equations can be rewritten in the form
$\nabla^a \xi^b + \nabla^b \xi^a  = 0$ (see our comments  below \eqref{B+}) and their explicit solution
reproduces the well-known $o(2,1)$ Killing vector parameterized by three integration constants representing
three $o(2,1)$ generators.

On the other hand, using the $\sigma_-$ cohomological reduction one finds out that
\eqref{transstab2} reduces to equations  \eqref{residualtr1}-\eqref{residualtr2}  or \eqref{vto}
on a scalar parameter $\xi^{(s)}$ at  $s = 1,2,..., \infty$ subjected to the  Bianchi identities \eqref{Bianchi3}.
In
this case the stability  transformations describe trivial
"improvement" transformations of the respective spin-$s$ conserved
currents. Contrary  to the general "improvement" transformations that
are invariance transformations of the conservation condition, the trivial "improvements"
do not change the conserved current itself. It seems that there is
no any "background conserved current" similar to the background spacetime, so that  an interpretation of
trivial "improvements" remains unclear.

\section{Summary of the metric-like  formulation}
\label{Sec:summaryMETR}

\subsection{Metric-like equations of motion}

Below we list the  metric-like equations following from the $\sigma_\pm$ cohomological reductions
of the original spin $s>1$ unfolded equation system  \eqref{2eq} and \eqref{1eq} analyzed in Sections \bref{sec:offshell}
and \bref{sec:0forms}.

\begin{itemize}

\item  \textbf{$\sigma_+$ - reduction}


\be
\label{1}
\hspace{0mm}\text{1-form sector:}  \hspace{1cm}   \big(\Box_{_{\ads}}   -s(s-1) \Lambda\big)\varphi = C\;,
\qquad \varphi^{a_1 ...\, a_s} = 0
\ee

\be
\label{2}
\ba{l}
\text{0-form sector:}  \hspace{1cm}
\epsilon_{mn_1}\nabla^m \nabla_{n_2} \cdots \nabla_{n_{s-1}}\varphi^{n_1 \cdots\, n_{s-1}}  = B_+
\\
\\
\hspace{33mm}\nabla_{(a_1} \varphi_{a_2 ... a_{s})} -\dps \frac{1}{s-1} \;\eta_{(a_1a_2} \nabla^c\, \varphi_{a_3 ... a_{s-1})c}
= B_{a_1... a_{s}}^{+}
\ea
\ee

\item  \textbf{$\sigma_-$ - reduction}

\be
\label{3}
\hspace{-25mm} \text{1-form sector:}  \hspace{1cm}   \nabla^n \phi_{n a_1 ... a_{s-1}} = C_{a_1 ... a_{s-1}}\;,
\qquad
\phi = 0
\ee

\be
\label{4}
\ba{l}
\hspace{-35mm}\text{0-form sector:}  \hspace{1cm}  \big(\Box_{_{\ads}}   -s(s-1) \Lambda\big)\phi  = B^-
\\
\\
\big(\nabla_{a_1} \cdots \nabla_{a_{s}} \phi + ...  \big)=
B_{a_1... a_{s}}^{-}
\ea
\ee

\end{itemize}

\noindent
Recall that the metric-like equations in the one-form sector have been  obtained  using the higher spin gauge
\eqref{hsgc} in the $\sigma_+$ case, and the scalar gauge \eqref{hsgc-} in the  $\sigma_-$ case.
In particular, the above equations are supplemented  with the leftover  gauge transformations and the Bianchi identities
in the one-form and the zero-form sectors, respectively. Note also that the metric-like equations of motion
are of order $1, 2, s-1, s$ in derivatives.

\subsection{Dual metric-like higher spin actions}
\label{sec:mapping}

Let us consider  linearized frame-like action \eqref{actBFlin}
in the metric-like form. To this end, we represent the action in Lorentz basis
\be
\label{actLor}
S_0[\phi,\omega] = \sum_{k=0}^{s-1}\int_{\cM^2}  \phi_{a_1 ... a_{k}}R^{a_1 ... a_k}(\omega)\;,
\ee
where  $0$-form fields $\phi_{a_1 ... a_{k}}$ are given by \eqref{dectracelessPHI} and $2$-form field strength
$R^{a_1 ... a_k}(\omega)$ is expressed
via  $1$-form gauge fields $\omega^{a_1 ... a_k}$ \eqref{curva}. The corresponding equations of motion
are given by  \eqref{main} and \eqref{main3} with vanishing right-hand-sides.

The idea is to fix  Stueckelberg (shift) gauge symmetries  and eliminate  auxiliary fields using their own equations
of motion
substituting then the independent metric-like fields and the field strengths  back to the frame-like action
\eqref{actLor}.  In particular,
this is the way one shows the equivalence of the frame-like $o(2,1)$ BF theory with the original metric-like
Jackiw-Teitelboim model \cite{Fukuyama:1985gg}.

As we have already seen, a reduction  to the independent dynamical sector can be done in two equivalent ways associated
either to $\sigma_+$ or $\sigma_-$ cohomology. Moreover, when considering both one-form and zero-form
sectors simultaneously one has four equivalent reductions which we denote as $(\sigma_{\pm}, \sigma_{\pm})$ reduction,
where the first and second sigmas refer to corresponding  reduction in the one-form and zero-form sector, respectively.
However, at the action level one finds out that
there are only  two possible ways to perform a reduction to the metric-like form.
Equations obtained via  $(\sigma_-, \sigma_-)$ or $(\sigma_+, \sigma_+)$ reductions cannot be derived as
variational equations since the number  of the independent fields do not coincide with the
number of the equations of motion.

Equations obtained via the $(\sigma_+$, $\sigma_-)$   reduction can be derived as the Euler-Lagrange equations of motion
following from the action
\be
\label{act+-}
S^{+-}_0[\varphi, \varphi_{a_1 ... a_s}| \phi] = \int_{\cM^2} \phi \, R(\varphi, \varphi_{a_1 ... a_s})\;,
\ee
where  $R(\varphi, \varphi_{a_1 ... a_s})$ is the 2-from field strength of grade degree $k=0$  \eqref{constr2} expressed
in terms of the dynamical fields. The equations of motion of the theory \eqref{act+-} take the form
\eqref{1} and \eqref{4} (using the higher spin gauge).  In particular, the linearized action and equations
of the Jackiw-Teitelboim model  follow from  \eqref{act+-} at $s=2$.

Analogously, equations obtained via the $(\sigma_-, \sigma_+)$ reduction follow  from the other
action
\be
\label{act-+}
S^{-+}_0[\phi, \phi_{a_1 ... a_s}| \varphi_{a_1 ... a_{s-1}}] =
\int_{\cM^2}  \varphi_{a_1 ... a_{s-1}}\, R^{a_1 ... a_{s-1}}(\phi, \phi_{a_1 ... a_s})\;,
\ee
where $R^{a_1 ... a_{s-1}}(\phi, \phi_{a_1 ... a_s})$ is the 2-form field strength of  grade degree
$k=s-1$
\eqref{constr1} expressed in terms of the dynamical fields. The  equations of motion
of the theory \eqref{act-+} take the form  \eqref{2} and \eqref{3} (using the scalar gauge).

The form of actions \eqref{act+-} and \eqref{act-+}  can be explained
by resorting to  the cross-duality \eqref{cohduality} exhibited by
the cohomology groups $H^{(m)}(\sigma_+)$ and $ H^{(n)}(\sigma_-)$
that gives, in particular, $H^{(2)}(\sigma_\pm) \approx
H^{(0)}(\sigma_\mp)$. To this end,  one employs inner product
\eqref{inner} on the space $\cG_s$ and reformulates action
\eqref{actLor} as $S_0[\phi,\omega] = \dps \int_{\cM^2} \left<
\phi| R\right >$, where $\phi,\omega, R \in \cG_s$. Then,  eliminating  the auxiliary fields via their own equations
of motion one finds that fields of the metric-like formulation  are elements of the cohomology,
$0$-forms $\left< \phi| \right. \in
H^{(0)}(\sigma_\pm)$ and reduced $2$-forms $\left. |R \right> \in
H^{(2)}(\sigma_\mp)$. After that, using the properties \eqref{pro1},
\eqref{pro2} along with  the above  cross-duality relation one
arrives at the two actions considered  above.

On the other hand, both types of the cohomological reductions describe the same
dynamical system. It suggests there exists a duality  mapping between two
linear theories given by \eqref{act+-} and \eqref{act-+}. It would
be interesting to provide an exact definition of such a mapping
originated from  the cohomology cross-duality and to study its
properties and implications beyond the linear approximation.

\subsection{The model interpretation}

The equations of motion in the one-form sector have been
previously interpreted as describing topological maximal depth
partially-massless higher spin fields on the   $\ads$ background
\cite{Alkalaev:2013fsa}. It should be noted that such an
interpretation follows from $(\sigma_+$, $\sigma_-)$ - reduction
described by action \eqref{act+-}.

In this case, the  equations of motion in both zero-form and
one-form sectors (in the gauge fixed form) are given by the same
Klein-Gordon equation $\big(\Box_{_{\ads}}   -s(s-1)
\Lambda\big)\varphi = 0$ and $\big(\Box_{_{\ads}}   -s(s-1)
\Lambda\big)\phi = 0$ for two scalars $\varphi$  and $\phi$. The
general solution depends on two arbitrary functions of spacetime
coordinates so that it can be interpreted as left and right waves.
However, there are gauge symmetry in the one-form sector and
additional tensor constraint along with the Bianchi identities in
the zero-form sector that eventually eliminate the functional
freedom leaving no local modes (only a finitely many integration constants).
The absence of propagating degrees of freedom
leaves enough room for interpretation of the equations of motion
under consideration. We set that fields in the one-form sector are
gauge fields, while those in the zero-form sector are  dilaton
fields, both topological.

The spectrum of the model can be interpreted as follows. The BF
higher spin theory  given by action \eqref{act+-} describes:
(one-form sector) topological $s=1$ massless Maxwell field and
$s=2$ graviton field along with increasing spin $s =3,4,...$ partially-massless
gauge fields of the maximal depth; (zero-form sector)
topological dilaton  fields with increasing masses $m^2_s = -s(s-1)
\Lambda$. In this form action \eqref{act+-} can be
treated  as a higher spin gauge-dilaton extension of the  original (linearized)
Jackiw-Teitelboim dilaton gravity model.

\section{ The higher spin algebras in two dimensions}
\label{sec:higher spin algebra}

To formulate a non-linear BF higher spin theory the fields should be represented as
connections of some (in)finite Lie algebra. In the case of  finitely many fields a higher spin algebra
can be identified with $sl(N, \mathbb{R})$ algebra provided that its basis elements are represented as
\be
\label{sln}
T_{A_1}\, \oplus\, T_{A_1A_2}\, \oplus\; \cdots \; \oplus\, T_{A_1 ... A_{N-1}}\;,
\ee
where generators $T_{A_1 ... A_k}$ are rank-$k$ totally symmetric and traceless $sl(2, \mathbb{R})$ algebra
tensors \cite{Bais:1990bs,Henneaux:2010xg,Campoleoni:2010zq}.
Gauging algebra \eqref{sln} yields a finite collection of $0$-form and $1$-form fields of the type \eqref{gaugefield}.
A natural  infinite-dimensional generalization of \eqref{sln} should have the following structure
\be
\label{infsln}
\bigoplus_{s=1}^\infty\, \bigoplus_{l_s}\, T^{(l_s)}_{A_1 ... A_{s-1}}\;,
\ee
where the numbers $l_s$ are multiplicities of  spin-$s$ basis elements. Note that \eqref{infsln}
contains also  infinitely many copies of $gl(1,\mathbb{R})$ generator $T$ corresponding
to the spin-$1$ Abelian connection.

A convenient way to realize higher spin algebras with generators
$T_{A_1 ... A_{s-1}}$ \eqref{infsln} is to represent them
as homogeneous polynomials of degree-$(s-1)$ in auxiliary
vector variables. It is remarkable that such a vector realization
can be obtained  using $d$-dimensional oscillator approach based on the $o(2,d-1)-sp(2)$ Howe duality
proposed  by Vasiliev \cite{Vasiliev:2003ev,Vasiliev:2004cm}. In
what follows, we use the  $o(2,1)-sp(2)$ Howe duality to describe
the one-parametric family of $2d$ higher spin algebras $\Hsalg$
originally introduced by Feigin as quotients of the universal
enveloping algebra $\cU(sl(2))$ \cite{Feigin}, and by Vasiliev  as
the enveloping algebra of the Wigner deformed oscillator algebra
\cite{Vasiliev:1989qh}.

\subsection{Oscillator approach}

Following the original papers \cite{Vasiliev:2003ev,Vasiliev:2004cm}, we consider auxiliary doublet variables $Y^A_\alpha$, with $sp(2)$ vector
index $\alpha$ and $o(2,M)$ vector index $A$, \footnote{In this section symplectic  indices
$\alpha,\beta,\gamma,... = 1,2$, vector indices $A,B,C ... = 0,..., M+1$, the $o(2,M)$ invariant metric
is $\eta_{AB} = (+-....-+)$, symplectic indices are raised and lowered with the $sp(2)$ invariant metric
$\epsilon_{\alpha\beta} =  - \epsilon_{\beta\alpha}$.  } and consider polynomials expanded in
the auxiliary variables as follows
\be
\label{polynomial}
F(Y) = \sum_{k=0}^\infty F_{A_1 ... A_k}^{\alpha_1 ... \alpha_k} Y^{A_1}_{\alpha_1} ... Y^{A_k}_{\alpha_k}
 =  \sum_{m,n = 0}^\infty F_{A_1 ... A_m | B_1 ... B_n} Y^{A_1}_1 \cdots Y^{A_m}_1\, Y^{B_1}_2 \cdots Y^{B_n}_2\;,
\ee
where expansion coefficients are totally symmetric in both groups of indices.

Define now the Weyl star-product
\be
\label{weyl}
(F*G)(Y) = \frac{1}{\pi^{2M}}\int dS dT \,F(Y+S)\,G(Y+T)\,\exp(-2S_\alpha^A T^\alpha_A)\;.
\ee
It follows that the auxiliary variables satisfy the following commutation relations
$\big[Y^A_\alpha, Y^B_\beta\big]_*  = \epsilon_{\alpha\beta}\eta^{AB}$. A space of polynomials
\eqref{polynomial} endowed with the star-product \eqref{weyl} is  the
Weyl algebra $\cA_{M+2}$.

The algebra $\cA_{M+2}$ is a bi-module over $o(2,M)$ and $sp(2)$ algebras. Their basis elements
are realized as bilinear combinations of the auxiliary variables
\be
\label{basis}
T^{AB} = \frac{1}{2}\epsilon^{\alpha\beta}\big\{ Y_\alpha^A, Y_\beta^B\big\}_*\;,
\qquad
t_{\alpha\beta} = \frac{1}{2}\eta_{AB}\big\{ Y_\alpha^A, Y_\beta^B\big\}_*\;.
\ee
Bilinears  $T^{AB}$ and $t_{\alpha\beta}$ commute, $[T^{AB},t_{\alpha\beta}]_* = 0$. Moreover,
the two algebras form  a Howe dual pair $o(2,M)-sp(2)$ \cite{Howe}. It follows that
$sp(2)$ highest weight conditions imposed on elements of $\cA_{M+2}$ single out particular
finite-dimensional $o(2,M)$ irreducible representations (see Section \bref{sec:S3} for more details).

Using  \eqref{basis} one finds that quadratic Casimir operators $C_2 = \half\, T_{AB}*T^{AB}$ of $o(2,M)$
algebra and $c_2 = \half\, t_{\alpha\beta}*t^{\alpha\beta}$ of $sp(2)$ algebra are related as
\be
\label{casimirs}
C_2 = \frac{1}{4}(M^2-4)+ c_2\;.
\ee

Higher spin algebras considered below are various quotients of the $*$-product algebra
 $\cS_{M+2} \subset \cA_{M+2}$ of all polynomials spanned  by $sp(2)$ invariant elements
\be
\label{spINV}
\big[t_{\alpha\beta}, F(Y)\big]_*=0\;.
\ee
Endowing the associative algebra $\cS_{M+2}$
with the commutator $[F,G]_*$, where $F,G \in \cS_{M+2}$ one obtains  the Lie algebra denoted as
$hc(1|2\hspace{-1mm}:\hspace{-1mm}[M,2])$ \cite{Vasiliev:2004cm}.
\footnote{In what follows, by a slight abuse of notation, we denote associative
algebras and Lie algebras obtained by taking the commutators with respect to the associative product by the same symbols.}

In general, associative algebra $\cS_{M+2}$ (as well as Lie algebra $hc(1|2\hspace{-1mm}:\hspace{-1mm}[M,2])$) contains various two-sided ideals $\cI$. For instance, there exists the maximal
ideal spanned by elements
\be
\label{ideal}
\cI_1 = \big\{\,g(Y) = t_{\alpha\beta}*g^{\alpha\beta}(Y)\,\big\}\;,
\qquad
\big[t_{\alpha\beta}, g^{\gamma\rho}\big]_* = \delta_\beta^\gamma \,g_\alpha{}^\rho + \text{3 terms}\;,
\ee
where $g^{\alpha\beta}(Y)$ is an arbitrary polynomial transforming as an $sp(2)$ symmetric tensor.
Using  ideals $\cI$ one  defines quotient algebras $\cH = \cS_{M+2}/\cI$.
So, factoring out the maximal ideal \eqref{ideal} gives rise to
associative  algebra $\cS_{M+2}/\cI_1$. A particular real form  of the
respective Lie algebra $hc(1|2\hspace{-1mm}:\hspace{-1mm}[M,2])/\cI_1$ is denoted as
$hu(1|2\hspace{-1mm}:\hspace{-1mm}[M,2])$ \cite{Vasiliev:2004cm}.
It is singled out by reality conditions
\be
\label{reality}
\big(F(Y)\big)^{\dagger} =  - F(Y)\;,
\ee
where the  involution $\dagger$ of the complex algebra $\cS_{M+2}$ is  defined as $(Y^A_\alpha)^\dagger = Y^A_\alpha$  and $\big(aF(Y)\big)^{\dagger}  = \bar a\big(F(Y)\big)^{\dagger}$,
where $a\in \mathbb{C}$, and the bar stands for the  complex
conjugation. Gauging $hu(1|2\hspace{-1mm}:\hspace{-1mm}[M,2])$ yields totally symmetric massless (Fronsdal)
fields of increasing spins $s = 1,2,..., \infty$.

In what follows, we explicitly consider  the case of $M=1$  and study quotient higher spin algebras corresponding
to different  ideals, including  the maximal one. We show that
$hc(1|2\hspace{-1mm}:\hspace{-1mm}[1,2])/\cI_1$
is a finite-dimensional  algebra. Therefore, in order to produce an infinite-dimensional
higher spin algebra one should  use non-maximal ideals. We identify two infinite families of
ideals that yield both finite- and infinite-dimensional quotient higher spin algebras.
 Our analysis  also applies  to the case of $M=2$, where the $AdS_3$ global symmetry algebra
 $o(2,2) \approx o(2,1)\oplus o(2,1)$, and each factor
can be considered by analogy with the case of $M=1$.

\subsection{Howe dual realization of $\cU(o(2,1))$}
\label{sec:S3}

Howe dual algebras $sp(2)$ and $o(M,2)$ act on $\cA_{M+2}$ so that
expansion coefficients of $F(Y)$ in the auxiliary variables \eqref{polynomial}
are both $sp(2)$ and $o(2,M)$ tensors. On the other hand,  the $sp(2)$ invariance  condition
\eqref{spINV} says that these tensors are of particular index symmetry type.
It follows that the resulting expansion coefficients of \eqref{polynomial} are $o(2,M)$  traceful tensors
with index symmetry described by rectangular two-row Young diagrams
\be
\label{rectangle}
F_{A_1 ... A_m,\, B_1 ... B_m}\;:\qquad \;\; F_{(A_1 ... A_m,\, B_1) B_2 ... B_m} \equiv 0\;.
\ee

In the $M=1$ case, any $o(2,1)$ traceful two-row rectangular  tensor  \eqref{rectangle} can be decomposed into one-row tensors
because any traceless $o(2,1)$ tensor with indices described by two-row Young diagram with more than one cell  in
the second row vanishes identically, while those with a single cell in the second row are dualized using
the Levi-Civita tensor, see \eqref{Hodge}.

It follows that a linear space of  the  algebra $\cS_3$ spanned by $sp(2)$ singlets  \eqref{spINV} can be represented as
an infinite collection of one-row traceless Young diagrams. Indeed, let $T_m$ denote a spin-$m$ $o(2,1)$ irrep given by a
totally symmetric traceless $o(2,1)$ tensor. Then, one can show that a linear
space of $\cS_3$  as o(2,1) module is decomposed in a direct sum
\be
\label{plot1}
\cS_3 = \bigoplus^\infty_{m=0}\,\bigoplus^\infty_{l=1}  \,T_m^{(l)}\;,
\ee
where a superscript $l$ stands for multiplicity, cf. \eqref{infsln}.
Elements of linear space \eqref{plot1} can be depicted on the following plot:

\be
\label{plot}
\ba{ccccccccccc}
\begin{picture}(-50,12)(0,10)
\put(-110,11){$\bullet $}
\end{picture}
&
\begin{picture}(12,12)(78,4)
{\linethickness{0.210mm}
\put(00,11){\line(1,0){10}} 
\put(00,03){\line(1,0){10}} 
\put(00,03){\line(0,1){8}} 
\put(10,03){\line(0,1){8}} 
}
\end{picture}
&
\begin{picture}(12,12)(80,4)
{\linethickness{0.210mm}
\put(00,11){\line(1,0){20}} 
\put(00,03){\line(1,0){20}} 
\put(00,03){\line(0,1){8}} 
\put(10,03){\line(0,1){8}} 
\put(20,03){\line(0,1){8}}
}
\end{picture}
&
\begin{picture}(12,12)(72,4)
{\linethickness{0.210mm}
\put(00,11){\line(1,0){30}} 
\put(00,03){\line(1,0){30}} 
\put(00,03){\line(0,1){8}} 
\put(10,03){\line(0,1){8}} 
\put(20,03){\line(0,1){8}}
\put(30,03){\line(0,1){8}}
}
\end{picture}
&
\begin{picture}(12,12)(54,4)
{\linethickness{0.210mm}
\put(00,11){\line(1,0){40}} 
\put(00,03){\line(1,0){40}} 
\put(00,03){\line(0,1){8}} 
\put(10,03){\line(0,1){8}} 
\put(20,03){\line(0,1){8}}
\put(30,03){\line(0,1){8}}
\put(40,03){\line(0,1){8}}
}
\end{picture}
&
\begin{picture}(12,12)(26,4)
{\linethickness{0.210mm}
\put(00,11){\line(1,0){50}} 
\put(00,03){\line(1,0){50}} 
\put(00,03){\line(0,1){8}} 
\put(10,03){\line(0,1){8}} 
\put(20,03){\line(0,1){8}}
\put(30,03){\line(0,1){8}}
\put(40,03){\line(0,1){8}}
\put(50,03){\line(0,1){8}}
}
\end{picture}
&
\begin{picture}(12,12)(-13,4)
{\linethickness{0.210mm}
\put(00,11){\line(1,0){60}} 
\put(00,03){\line(1,0){60}} 
\put(00,03){\line(0,1){8}} 
\put(10,03){\line(0,1){8}} 
\put(20,03){\line(0,1){8}}
\put(30,03){\line(0,1){8}}
\put(40,03){\line(0,1){8}}
\put(50,03){\line(0,1){8}}
\put(60,03){\line(0,1){8}}

}
\end{picture}
&
\begin{picture}(12,12)(-63,4)
{\linethickness{0.210mm}
\put(00,11){\line(1,0){70}} 
\put(00,03){\line(1,0){70}} 
\put(00,03){\line(0,1){8}} 
\put(10,03){\line(0,1){8}} 
\put(20,03){\line(0,1){8}}
\put(30,03){\line(0,1){8}}
\put(40,03){\line(0,1){8}}
\put(50,03){\line(0,1){8}}
\put(60,03){\line(0,1){8}}
\put(70,03){\line(0,1){8}}

}
\end{picture}
&
\begin{picture}(12,12)(-120,10)
{\linethickness{0.210mm}
\put(00,11){$\cdots$} 
}
\end{picture}
&
&
\\
&
&
&
&
\begin{picture}(12,12)(123,10)
{\linethickness{0.210mm}
\put(00,11){$\bullet$} 
}
\end{picture}
&
\begin{picture}(12,12)(116,4)
{\linethickness{0.210mm}
\put(00,11){\line(1,0){10}} 
\put(00,03){\line(1,0){10}} 
\put(00,03){\line(0,1){8}} 
\put(10,03){\line(0,1){8}} 
}
\end{picture}
&
\begin{picture}(12,12)(98,4)
{\linethickness{0.210mm}
\put(00,11){\line(1,0){20}} 
\put(00,03){\line(1,0){20}} 
\put(00,03){\line(0,1){8}} 
\put(10,03){\line(0,1){8}} 
\put(20,03){\line(0,1){8}}
}
\end{picture}
&
\begin{picture}(12,12)(70,4)
{\linethickness{0.210mm}
\put(00,11){\line(1,0){30}} 
\put(00,03){\line(1,0){30}} 
\put(00,03){\line(0,1){8}} 
\put(10,03){\line(0,1){8}} 
\put(20,03){\line(0,1){8}}
\put(30,03){\line(0,1){8}}
}
\end{picture}
&
\begin{picture}(12,12)(31,4)
{\linethickness{0.210mm}
\put(00,11){\line(1,0){40}} 
\put(00,03){\line(1,0){40}} 
\put(00,03){\line(0,1){8}} 
\put(10,03){\line(0,1){8}} 
\put(20,03){\line(0,1){8}}
\put(30,03){\line(0,1){8}}
\put(40,03){\line(0,1){8}}
}
\end{picture}
&
\begin{picture}(12,12)(-19,4)
{\linethickness{0.210mm}
\put(00,11){\line(1,0){50}} 
\put(00,03){\line(1,0){50}} 
\put(00,03){\line(0,1){8}} 
\put(10,03){\line(0,1){8}} 
\put(20,03){\line(0,1){8}}
\put(30,03){\line(0,1){8}}
\put(40,03){\line(0,1){8}}
\put(50,03){\line(0,1){8}}
}
\end{picture}
&
\begin{picture}(12,12)(-77,10)
{\linethickness{0.210mm}
\put(00,11){$\cdots$} 
}
\end{picture}
\\

&

&

&

&

&

\begin{picture}(12,12)(77,10)
{\linethickness{0.210mm}
\put(00,11){$\bullet$} 
}
\end{picture}
&
\begin{picture}(12,12)(48,4)
{\linethickness{0.210mm}
\put(00,11){\line(1,0){10}} 
\put(00,03){\line(1,0){10}} 
\put(00,03){\line(0,1){8}} 
\put(10,03){\line(0,1){8}} 
}
\end{picture}
&
\begin{picture}(12,12)(9,4)
{\linethickness{0.210mm}
\put(00,11){\line(1,0){20}} 
\put(00,03){\line(1,0){20}} 
\put(00,03){\line(0,1){8}} 
\put(10,03){\line(0,1){8}} 
\put(20,03){\line(0,1){8}}
}
\end{picture}
&
\begin{picture}(12,12)(-41,4)
{\linethickness{0.210mm}
\put(00,11){\line(1,0){30}} 
\put(00,03){\line(1,0){30}} 
\put(00,03){\line(0,1){8}} 
\put(10,03){\line(0,1){8}} 
\put(20,03){\line(0,1){8}}
\put(30,03){\line(0,1){8}}
}
\end{picture}
&
\begin{picture}(12,12)(-100,10)
{\linethickness{0.210mm}
\put(00,11){$\cdots$} 
}
\end{picture}
\\

&

&

&

&

&

&

&
\begin{picture}(12,12)(11,10)
{\linethickness{0.210mm}
\put(00,11){$\bullet$} 
}
\end{picture}
&
\begin{picture}(12,12)(-41,4)
{\linethickness{0.210mm}
\put(00,11){\line(1,0){10}} 
\put(00,03){\line(1,0){10}} 
\put(00,03){\line(0,1){8}} 
\put(10,03){\line(0,1){8}} 
}
\end{picture}
&
\begin{picture}(12,12)(-100,10)
{\linethickness{0.210mm}
\put(00,11){$\cdots$} 
}
\end{picture}

\ea
\ee

\noindent Here, irreps $T_k$ are depicted as length-$k$ Young diagrams,  dots  $ \bullet$ correspond to scalar components $T_0$.
Irreps $T_k$ resulted from decomposing a traceful two-row rectangle of length $m-1$
are disposed vertically, $k = 0,...,m$. Note that an
each line on the plot successively depicts all basis elements of $gl(N)$ algebra, where
$N = 1,2,3,...$.

The other way around,  traceless symmetric tensors can be rearranged as
 traces of a given totally symmetric traceful tensor.
It suggests that the linear  space can be described by traceful symmetric tensors
of all ranks from zero to infinity, each in a single copy. It can be equivalently seen by dualizing
traceful rectangular $o(2,1)$ diagrams \eqref{rectangle}. It follows that the linear space of $\cS_3$
can be represented as
\be
\label{plot111}
\cS_3 =  \bigoplus^\infty_{k=0} \,G_k\;,
\ee
where $G_k$ denotes a rank-$k$ symmetric traceful $o(2,1)$ tensor; it follows  that
$G_k = T_k \oplus T_{k-2} \oplus \cdots  \,$. On the plot \eqref{plot}  a tensor
$G_k$ corresponds to the $k$-th vertical column.

Let us now notice that when indices $A, B, ...$ run just three
values it is possible to introduce new variables
\be
\label{Tdual}
T_A = \epsilon_{ABC} \epsilon^{\alpha\beta} Y^B_\alpha Y^C_\beta\;,
\ee
which are in fact Hodge dualized $o(2,1)$ basis elements \eqref{basis}, and hence satisfy the commutation
relations $\big[T_A, T_B\big]_* =\epsilon_{ABC}T^C$.
One can show that any $sp(2)$ singlet $F(Y)$ can be equivalently rewritten as an arbitrary polynomial
$F(T)$.  Indeed,  the $sp(2)$ invariance condition \eqref{spdecom}  says  that
expansion coefficients of any $F \in \cS_3$ \eqref{polynomial}
have even numbers of $sp(2)$ and $o(2,1)$ vector indices, and can be represented as
\be
\label{FFF}
F_{A_1 ... A_{2m}}^{\alpha_1 ... \alpha_{2m}} = \epsilon^{\alpha_1\alpha_2} ...\, \epsilon^{\alpha_{2m-1}\alpha_{2m}}
F_{A_1A_2 | \,\cdots\, |A_{2m-1}A_{2m}}\,
\ee
where each group of two vector indices $|A_i A_{i+1}|$ is antisymmetric (see \cite{Alkalaev:2007bq} for more
details). Using the definition  \eqref{Tdual} along with \eqref{FFF} one finds that \eqref{polynomial} can be
completely rewritten as polynomials  of $o(2,1)$ bilinears  $T^{A}$ with totally symmetric expansion coefficients.
Note that $T^A$ are $sp(2)$ singlets. It follows that the space $\cS_3$ of $sp(2)$ singlets is now naturally
realized as functions of $sp(2)$ invariant variables. The action of Howe dual algebra $sp(2)$ becomes implicit.

In this way, we establish that the associative algebra $\cS_3$ of $sp(2)$ singlets and the universal enveloping algebra $\cU(o(2,1))$ are isomorphic,
\be
\label{iso}
\cS_3 \approx \cU(o(2,1))\;.
\ee

Note that the above consideration applies  to $\cS_{M+2}$ for any $M$.
However,  its basis elements are parameterized by $o(2,M)$ two-row
rectangle $o(2,M)$ diagrams \eqref{rectangle} so that $\cS_{M+2}$ cannot be
interpreted as the universal enveloping algebra $\cU(o(2,M))$. In
the  case of $M=1$ two-row rectangle diagrams  become arbitrary
one-row diagrams making  isomorphism  \eqref{iso} possible.

\vspace{-3mm}

\paragraph{Trace decomposition.} Subtracting $o(2,1)$ traces can be done systematically if one employs $sp(2)$ Howe dual algebra.
To this end, consider first $o(2,M)$ trace decompositions. From the definition of $sp(2)$ basis elements  $t_{\alpha\beta}$ \eqref{basis} it follows that all three possible
traces of a tensor with indices described by $o(2,M)$ two-row Young diagram can be collectively represented  as
three
independent $sp(2)$ generators. In particular,  any multiple trace of $F\in \cS_3$
is  to be proportional to the following combination \cite{Alkalaev:2007bq}
\be
\label{tracesp}
t_{\alpha\beta}\, \cdots\,  t_{\gamma\rho}\; c_2\, \cdots \, c_2\;.
\ee
Here, $sp(2)$ indices are assumed to be symmetrized. Totally antisymmetric combinations of $t_{\alpha\beta}$
produces powers of the $sp(2)$ Casimir element $c_2$.

By way of example consider  particular polynomial $F(Y) = F_{AB | CD} Y^A_1 Y^B_1 Y^C_2 Y^D_2$
subjected to the $sp(2)$ invariance condition \eqref{spINV}. It follows
that an  expansion coefficient $F_{AB,\, CD}$ is described by a "window" Young diagram
$\begin{picture}(13,12)(-1,1)
{\linethickness{0.210mm}
\put(00,10){\line(1,0){10}}
\put(00,05){\line(1,0){10}}
\put(00,00){\line(1,0){10}}
\put(00,00){\line(0,1){10}}
\put(05,00){\line(0,1){10}}
\put(10,00){\line(0,1){10}}
}
\end{picture}
$.
On the other hand, the expansion coefficient is traceful so that a
decomposition into traceless parts yields a linear combination
\be
\label{deco}
F_{AB,\, CD} = F^0_{AB,\, CD} + \eta_{AB} F^1_{CD} + \eta_{AB}\eta_{CD} F^2 + \ldots\;,
\ee
where the ellipsis denote proper symmetrization of indices, while $F^0_{AB,\, CD}$, $F^1_{AB}$, and $F^2$
are traceless components. Substituting the above decomposition into $F(Y)$ one finds  that
the second term is  proportional to $t_{\alpha\beta}$, while the third term  is
proportional to $c_2$, \textit{i.e.}, $F(Y) = F_0(Y)+ t_{\alpha\beta}F_1^{\alpha\beta}(Y) + c_2 F_2$.
For the case of $M=1$ the first term in decomposition \eqref{deco} identically vanishes, $F^0_{AB,\, CD} = 0$.
The second and the third terms correspond to $T_2$ and $T_0$ elements depicted in the third vertical column on  the plot \eqref{plot}.

It follows that a trace decomposition of any  $F(Y) \in \cS_3$ reads \cite{Alkalaev:2007bq}
\be
\label{spdecom}
F(Y) = F_0+ F_1(Y)+ \sum_{k,m=0}^\infty F_{(m)}^{\alpha_1 ... \alpha_{2k}}(Y)
\,t_{\alpha_1\alpha_2}  \cdots   t_{\alpha_{2k-1}\alpha_{2k}}\, (c_2)^m\;,
\ee
where $F_0$ and $F_1(Y)$ denote the  scalar and the vector components, while
$F_{(m)}^{\alpha_1 ... \alpha_{2k}}(Y)$ are totally symmetric $sp(2)$ rank-$2k$ tensors,  a subscript $m$ stands for a multiplicity.
Using the symmetry  property
$F^{...\alpha\beta...}_{...AB...} = F^{...\beta\alpha...}_{...BA...}$ one concludes that expansion coefficients in \eqref{spdecom}
are given by totally symmetric $o(2,1)$ traceless tensors. It is worth noting that analogous decomposition for elements of $\cS_{M+2}$ algebra is 3-parametric, while taking $M=1$ leaves only 2 parameters.
The absent branch corresponds to traceless two-row rectangular $o(2,M)$ Young diagrams. In the case $M=1$  this
branch reduces to the two first terms.

One concludes  that  the first line in  \eqref{plot} contains $T_k$ for $k\geq 2$ that appear as
coefficients in front of symmetrized combinations
$t_{(\alpha_1\alpha_2} * ... * t_{\alpha_{2k-1}\alpha_{2k})}$, while
subsequent lines necessarily contain powers of $c_2$. Any tensor on the plot \eqref{plot}
is proportional to particular combination \eqref{tracesp} except for the first two scalar $T_0$ and vector
$T_1$ representations.

\subsection{Quotient higher spin algebras}
\label{sec:quotients}

Algebra $\cS_3$ is not simple. In what follows, we consider
two  types of ideals $\cI \subset \cS_3$ along with  respective quotient algebras $\cS_3/\cI$ which we call vertical and horizontal ones according
to their graphical interpretation \eqref{plot} and trace decomposition \eqref{spdecom}.

For instance, factoring out the maximal ideal $\cI_1$ spanned  by elements \eqref{ideal} yields  the quotient
$\cH_1 = \cS_3/\cI_1$ spanned by a finitely many basis elements
\be
\label{QH1}
\cH_1 = T_0 \oplus T_1\;,
\ee
corresponding to  $gl(2,\mathbb{R}) \approx gl(1,\mathbb{R})\oplus sl(2,\mathbb{R})$ algebra. Indeed, using the trace decomposition
\eqref{spdecom} one notes that  all elements in \eqref{plot} save  for $T_0$ and $T_1$  are proportional to
$sp(2)$ generators $t_{\alpha\beta}$. It follows  that all such elements belong to the  ideal $\cI_1$ and therefore
are to be factored out.

\subsubsection{Horizontal  factorization}
\label{sec:horizont}

The maximal ideal is the first element in a family of two-sided ideals
\be
\label{ideal1}
\cI_k= \big\{\,T_{\alpha_1 ... \alpha_{2k}}* g^{\alpha_1  ... \alpha_{2k}}(Y)\,\big\}\;,
\qquad k \in \mathbb{N} \;,
\ee
where
\be
\label{PPP}
T_{\alpha_1 \alpha_2 ... \alpha_{2k}} = t_{(\alpha_1\alpha_2} * ... * t_{\alpha_{2k-1}\alpha_{2k})}\;,
\ee
and  $g^{\alpha_1  ... \alpha_{2k}}(Y)$ is a  rank-$2k$ symmetric $sp(2)$ tensor: $\big[t_{\gamma\rho}, g^{\alpha_1\alpha_2 ... }\big]_* = \delta_\rho^{\alpha_1} \,g_\gamma{}^{\alpha_2}
+ \ldots\; $, where the ellipses denotes all possible symmetrizations.
Using the associativity of the $*$-product, the $sp(2)$-invariance condition
\eqref{spINV}, and the following elementary properties
\be
\ba{l}
\dps
\big[t_{\gamma\rho}\,,\, g^{\gamma\rho \alpha_3  ... \alpha_{2k}}(Y)\big]_* = 0\;,
\\
\\
\dps
\big[T_{\alpha_1 ... \alpha_{2k}}\,,\, g^{\alpha_1  ... \alpha_{2k}}(Y)\big]_* = 0 \;,
\\
\\
\dps
\big[F(Y)\,,\, T_{\alpha_1 \alpha_2 ... \alpha_{2k}}\big]_* = 0\;, 
\ea
\ee
where $F(Y) \in \cS_3$,  one shows that $\cI_k \subset  \cS_3$  is a two-sided  ideal.
Note that  ideals \eqref{ideal1}
form an infinite flag sequence
\be
\label{flag}
\cI_1  \supset \cI_{2} \supset  \cdots \supset \cI_{k}\supset \cdots \;.
\ee

A  quotient algebra  $\cH_k = \cS_3/\cI_k$ is given by
\be
\label{carspa}
\cH_k  = \bigoplus_{m=0}^{2k-1} \, G_m\;.
\ee
cf. \eqref{plot111}. It is finite-dimensional and isomorphic  to a direct sum of
general linear algebras
\be
\label{horizontalgebra}
\cH_k \approx  gl(2, \mathbb{R}) \oplus  ... \oplus gl(2k-2, \mathbb{R}) \oplus gl(2k, \mathbb{R})\;.
\ee
To prove \eqref{horizontalgebra} one notes that factoring out elements proportional
to \eqref{PPP} for a given $k$ is equivalent to truncating the plot \eqref{plot}  starting from $(2k+1)$-th column.
The remaining elements form   \eqref{carspa}.

\subsubsection{Vertical  factorization}

Another type of ideals is given by a family
\be
\label{It}
\cI^t = \big\{\,I_t(c_2)* g(Y)\;, \;\;  \forall g \in \cS_3\,\big\}\;,
\ee
where $I_t(c_2)$ is a $t$-th order  $*$-product polynomial in the $sp(2)$ Casimir element $c_2$. Using the $sp(2)$
invariance condition \eqref{spINV} one shows that $\cI^t\subset \cS_3$ are two-sided ideals.
From \eqref{spdecom} and \eqref{plot} it follows that the resulting quotient algebra $\cH^t = \cS_3/\cI^t$ is given by
\be
\label{Ht}
\cH^t = \dps \bigoplus^\infty_{m=0}\,\bigoplus_{l=1}^{t} \,  \, T_m^{(l)}\;.
\ee

Any polynomial $I_t(c_2)$ can be decomposed into elementary monomials, so that
an ideal corresponding to $I_1 = c_2+\nu$, where $\nu$ is a constant parameter,
\be
\label{ideal2}
\cI^1_\nu = \big\{\,(c_2+\nu)* g(Y)\;, \;\; \forall g \in \cS_3\big\}\;,
\ee
is special. Taking  $t=1$ in \eqref{Ht} one arrives at the quotient algebra $\cH^1_\nu = \cS_3/\cI^1_\nu$ given by
\be
\label{Hu1}
\cH^1_\nu =  \dps \bigoplus^\infty_{m=0}\, \, T_m\;.
\ee

Recalling that  $\cS_3 \approx \cU(o(2,1))$  \eqref{iso} and using
the relation $c_2 = C_2 + \frac{3}{4}$ obtained by taking $M=1$ in formula \eqref{casimirs},
one finds that the above factorization is equivalent to factoring out elements proportional to $C_2 + \frac{3}{4}$
from the universal enveloping algebra $\cU(o(2,1))$.
In this way, we obtain that $\cH^1_\nu  = \cU(o(2,1))/\cI_{C_2 + \frac{3}{4}+\nu}$, and, therefore,
$\cH^1_\nu$ is isomorphic to the higher spin algebra $\Hsalg$ \cite{Feigin,Vasiliev:1989qh,Bergshoeff:1991dz}.
On the other hand, the algebra $\Hsalg$ is spanned  by  polynomials of two
spinor variables $q_\alpha$ and an idempotent element $K$
with commutation relations $[q_\alpha, q_\beta] = 2i \epsilon_{\alpha\beta}(1+ \nu K)$, $\;\{q_\alpha, K\} = 0$
\cite{Vasiliev:1989qh}.

Note that the two types of factorizations can be  visualized on the plot \eqref{plot}. The horizontal
factorization corresponds to truncating the plot horizontally  starting from $(2k+1)$-th  column.
The vertical  factorization  corresponds to truncating the plot vertically  starting from $t$-th row.

\subsubsection{Double factorizations}
\label{sec:doublefack}

For particular integer  $\nu$ algebra $\cH^1_\nu$ \eqref{Hu1} contains an additional (infinite-dimensional) ideal. The corresponding quotient
is a finite-dimensional general linear algebra  \cite{Feigin,Vasiliev:1989qh}. Using the $o(2,1)-sp(2)$ Howe duality
this can be seen as follows.

For a given $\nu$, all other ideals $\cI^1_\mu$ for $\mu\neq \nu$
and ideals $\cI_k$ \eqref{ideal1} for any $k$  in the quotient $\cS_3/\cI^1_{\nu}$  become
the trivial ideal which is the entire quotient itself.

Indeed, factoring out $\cI_{\nu}^1$ one obtains that in the quotient
algebra $\cH^1_{\nu}$ the $sp(2)$ Casimir element takes a particular value $c_2 = -\nu$. Consider now ideal
$\cI_\mu \subset \cS_3$ with parameter $\mu \neq \nu$. Using definition \eqref{ideal2} one shows that elements of
$\cI_\mu$ restricted to quotient $\cH^1_{\nu}$ are of  the form $(\mu-\nu) g$, where $g \in \cH^1_{\nu}$.
As a result, $\cI_\mu^1 \approx \cH^1_{\nu}$ for $\mu\neq \nu$, and $\cI^1_{\mu} \approx \varnothing$ for $\mu=\nu$,
so that the ideal becomes trivial.

The same reasoning applies to another type of ideals $\cI_k$ restricted to the quotient algebra
$\cH^1_{\nu}$. To this end, taking in \eqref{ideal1} elements
$g^{\alpha_1  ... \alpha_{2k}}(Y) = T^{\alpha_1  ... \alpha_{2k}}(Y)* g(Y)$, where  $\forall g(Y) \in \cS_3$,
and using the formula
\be
\label{P2}
T_{\alpha_1 \alpha_2 ... \alpha_{2k}} * T^{\alpha_1 \alpha_2 ... \alpha_{2k}}  = \tau_k \prod_{m=0}^{k-1}
* \,(c_2 + \alpha_m)\;,
\qquad
\alpha_m = m(2m+1)\;,
\ee
where $\tau_k$ is some non-vanishing  normalization coefficient, one  shows  that $\cI_k$ contains elements
$g(Y)*\prod_{m=0}^{k-1} * \,(c_2 + \alpha_m)$, where $\alpha_m = m(2m+1)$. Substituting the quotient value
$c_2 = - \nu$
one finds that $\cI_k$ contains elements of the form $g(Y) \prod_{m=0}^{k-1} \,(\alpha_m - \nu_0)$, where $g(Y)\in \cH^1_{\nu_0}$.
For general values $\nu$ the appearance of these elements implies that the ideal $\cI_k$ is trivial,
\textit{i.e.}, $\cI_k \approx \cH_{\nu}$.

However, for particular integer values
\be
\label{u0}
\nu_0=(k-1)(2k-1)\;,
\qquad  k \in \mathbb{N}\;,
\ee
one finds that the  ideal $\cI_k$ restricted to $\cH^1_{\nu_0}$ is non-trivial, and, therefore, can be factored out.
Indeed, ideal $\cI_k$ restricted to $\cH^1_{\nu_0}$ does not contain any powers of the $sp(2)$ Casimir element  since
$c_2 = -\nu_0$.
On the other hand, it contains combinations $T_{\alpha_1 \alpha_2 ... \alpha_{2l}}$  for $l \geq k$ only,
cf. \eqref{PPP} and \eqref{flag}.
Since the horizontal  factorization yields a finite-dimensional quotient, we conclude that the result
of such a double factorization is finite-dimensional as well: examining  the plot \eqref{plot} one finds out that
basis  elements of the double factorization span a general linear  algebra,
\be
\label{feigin}
\cH^1_{\nu_0}/\cI_k \approx  gl(2k, \mathbb{R})\;.
\ee
Note  that the rank of the algebra \eqref{feigin}  is even.
In Conclusions \bref{sec:concl} we discuss how to take account of odd values.

Finally, one can use a combination of the  two types of ideals in a single  factorization. For instance, consider
a composite  two-sided ideal
$\cI^p_1 = \big\{\,t_{\alpha\beta}*I_p(c_2)*  g^{\alpha\beta}(Y)\,\big\}$ provided that a $sp(2)$ symmetric tensor
$g^{\alpha\beta}$ is not proportional to $t^{\alpha\beta}$, and $I(c_2)$ is some $p$-th order  polynomial in $c_2$.
The resulting quotient algebra is given by
\be
\label{Hp}
\cH^p_1 = \dps  \Big[T_0 \oplus T_1\Big] \oplus \Big[\bigoplus^\infty_{m=0}\,\bigoplus_{l=1}^{p}\,  \, T_m^{(l)}\Big]\;.
\ee

\subsection{Factorization via (quasi-)projectors}
\label{sec:factor}

To describe quotients of  algebra $\cS_3$  explicitly  one employs
the projecting technique elaborated in
\cite{Vasiliev:2001wa,Vasiliev:2004cm}. \footnote{The projecting technique
was also discussed  in Refs.  \cite{Sezgin:2001zs,Sezgin:2001ij,Alkalaev:2002rq,Sagnotti:2005ns,Alkalaev:2010af}.} Given a quotient $\cH$ of algebra $\cS_3$
with respect to some ideal $\cI$ one introduces a quasi-projector
$\Delta$ satisfying the basic property
\be
\label{defpro}
\Delta * h = h*\Delta = 0\;, \qquad \forall\, h \in \cI \;.
\ee
Then, it follows that elements of  quotient $\cH = \cS_3/ \cI  $ can be parameterized as follows
\be
\cH= \big\{ g \in \cH: \;\; g = \Delta * F\;, \;\forall \,F\in \cS_3\big\}\;.
\ee
An educated guess is to consider the  following ansatz
\be
\label{deltaAN}
\Delta = \Delta(z)\;,
\qquad
z = Y_\alpha{}_A Y_\beta^A Y^\alpha_B Y^\beta{}^B\;.
\ee
Note that $z = 2c_2 - 9/2$, where $c_2$ is $sp(2)$ Casimir operator.  Variable $z$ is invariant with respect to both $sp(2)-o(2,1)$ Howe dual algebras, $[t_{\alpha\beta},z]_* = 0$ and
$[T^{A},z]_* = 0$. In particular,
\be
\label{cyclDelta}
\forall\, F \in \cS_3\,: \qquad \Delta *F = F*\Delta \;.
\ee

In Appendix \bref{sec:appendixB} we explicitly analyze the projecting conditions \eqref{defpro} imposed on
 $\Delta(z)$ \eqref{deltaAN}.
We show that the horizontal projecting condition is given by an ordinary $2k$-th order
differential equation for function  $\Delta_k(z)$.  The vertical projecting condition
is an ordinary $4$-th order differential equation for function $\Delta_\nu(z)$.
In both cases the searched-for solutions have the form of the series
$\Delta(z) = \kappa_0z^\alpha + \kappa_1 z^{\alpha+1}+ \kappa_2 z^{\alpha+2} + \cdots$, for some degree
$\alpha \geq 0$ and fixed coefficients $\kappa_i$ depending on either $k$ or $\nu$. Also, we analyze
solutions with  parameter $\nu$ taking  particular values \eqref{u0}.

\section{Non-linear higher spin BF  action}
\label{sec:action}

As a starting point, we formulate a non-linear higher spin theory in two dimensions as BF theory with gauge fields
taking values in the adjoint representation of the infinite-dimensional Lie algebra
$hc(1|2\hspace{-1mm}:\hspace{-1mm}[1,2])$ explicitly discussed in Section \bref{sec:S3}. After that, using the factorization
procedure of Section \bref{sec:factor} we describe reduced theories with fields taking values in the
quotient higher spin Lie  algebras.

The fields of the theory are
$0$-forms and $1$-forms taking values in $hc(1|2\hspace{-1mm}:\hspace{-1mm}[1,2])$ algebra
\be
\label{01}
\ba{c}
\dps
\Psi(Y|x)\;,
\qquad
W(Y|x) = dx^m  W_m(Y|x)\;.

\ea
\ee
From  \eqref{plot1} it follows that the expansion coefficients in the auxiliary variables  of  \eqref{01}  are
$0$-form and $1$-form fields taking values in totally symmetric traceless $o(2,1)$
representations of any rank. Each independent field  enters in infinitely many copies,
cf. \eqref{infsln}. We assume that fields \eqref{01} satisfy the reality conditions
\be
\Psi^\dagger(Y) = - \Psi(Y)\;,
\qquad
W^\dagger(Y) = - W(Y)\;,
\ee
where the conjugation  $\dagger$ is defined by \eqref{reality}.

The higher spin curvature associated to $1$-form gauge fields \eqref{01} is defined as
\be
\label{fullcurvature}
\cR(Y|x) = dx^m  dx^n \cR_{mn}(Y|x) = d W(Y|x) + W(Y|x) * W(Y|x)\;,
\ee
while the infinitesimal gauge transformations are
\be
\label{HSgaugetr}
\delta_\varepsilon W  = D \varepsilon\;,
\qquad
\delta_\varepsilon \Psi = [\Psi, \varepsilon]_*\;,
\qquad
\delta_\varepsilon \cR =  [\cR, \varepsilon]_*\;,
\ee
where $\varepsilon = \varepsilon(Y|x)$ is $0$-form gauge parameter taking values in the algebra $hc(1|2\hspace{-1mm}:\hspace{-1mm}[1,2])$, and
\be
\label{derivative}
D F = dF + [W,F]_*\;, \qquad d = dx^m \frac{\d}{\d x^m}\;,
\ee
is the gauge covariant derivative.

Consider now an invariant bilinear form on the higher spin algebra needed to build a BF action.
To this end,  define a trace of any element $F(Y) \in hc(1|2\hspace{-1mm}:\hspace{-1mm}[1,2])$
as follows \cite{vasiliev_oscillator}
\be
\label{trace}
\mathrm{Tr}(F(Y)) = F(0)\;.
\ee
The trace satisfies the cyclic property
\be
\mathrm{Tr}\big(F*G - G*F\big) = 0\;,
\qquad
\forall F, G \in hc(1|2\hspace{-1mm}:\hspace{-1mm}[1,2])\;,
\ee
that can be directly shown  using the definition \eqref{weyl} and the property that $F$ is even function,
$F(Y) = F(-Y)$. It follows that the algebra $hc(1|2\hspace{-1mm}:\hspace{-1mm}[1,2])$ can be
endowed with the following invariant bilinear form
\be
\label{invform}
\langle F, G  \rangle = \mathrm{Tr}(F *G)\;,
\ee
which is symmetric $\langle F, G  \rangle = \langle G,F  \rangle$ and  invariant
$\big\langle [F,G]_*, H  \big\rangle =  \big\langle G, [H, F]_*  \big\rangle$. From \eqref{weyl}
it follows that the invariant form has an integral representation useful in practice.

Using the invariant  bilinear form \eqref{invform} one defines the higher spin BF  action as
\be
\label{action}
S[\Psi, W] = g \int_{\cM^2} \mathrm{Tr}\,(\Psi * \cR) 
\ee
where $g$ is a dimensionless coupling constant. The above action can be invariantly extended by adding potentials
which are linear combinations of Casimir  polynomials $\kappa_i I_i(\Psi)$ on the algebra, where
$\kappa_i$ are coupling constants.

The equations of motion obtained by varying with respect to  $W_m(Y|x)$ and $\Psi(Y|x)$ are
\be
\label{covcon}
\cR_{mn}(Y|x)  = 0\;,
\ee
and
\be
\label{curveq}
D_m \Psi(Y|x) = 0\;,
\ee
where the gauge  covariant derivative $D_m$ is given by \eqref{derivative}.
The equation \eqref{covcon} is the covariance constancy condition
involving  both fields $\Psi$ and $W_m$, while equation the
\eqref{curveq} is the zero-curvature condition involving  fields
$W_m$ only. It follows that the gauge sector of the theory can be
analyzed independently.   Adding  invariant  potentials to the
action results in that the curvature acquires non-vanishing right-hand-side.
For instance, additional   terms proportional to the second-order
invariant  operator  $I_2 = \mathrm{Tr}(\Psi*\Psi)$ yields the
deformation \eqref{augment} discussed earlier within  the linearized
theory.

By construction, the higher spin BF  action is invariant under the gauge symmetry transformations \eqref{HSgaugetr}.
On the other hand, the theory is manifestly diffeomorphism invariant as it is formulated via differential forms, while
containing no metric
tensor. The diffeomorphism transformations of fields \eqref{01} are given by the respective Lie derivatives
\be
\delta_\xi \Psi = \xi^m \d_m \Psi\;,
\qquad
\delta_\xi W_n = \xi^m \d_m W_n+ \d_n \xi^m W_m\;,
\ee
that can be represented as follows
\be
\ba{c}
\dps
\delta_\xi \Psi =  \big[\Psi, \xi^m W_m\big]_* +\xi^m D_m \Psi  \;,
\qquad
\delta_\xi W_n = D_m\big(\xi^n W_n\big) + \xi^n \cR_{nm}\;.
\ea
\ee
The terms proportional to the field equations
represent the trivial invariance transformations vanishing
on the mass-shell.
Indeed, given  any  action $S[\phi_i]$ depending on fields $\phi_i$, $i=1,2,3,...$ one has a trivial invariance
transformation  $\delta \phi_i = M_{ij}\, \delta S/\delta \phi_j$, where the parameter matrix is
antisymmetric $M_{ij} = -M_{ij}$. Symmetries which differ by these trivial terms are equivalent.
In our case, 0-form $\Psi$ and 1-form $W$ are identified with $\phi_1$ and $\phi_2$.
It follows that modulo the trivial transformations the  diffeomorphisms
are just a particular gauge transformation with a field-dependent gauge parameter, and, therefore,  can be disregarded as independent symmetries.
\footnote{In particular, for the spin
$s=1$ two components of the diffeomorphism parameter $\xi^n(x)$
combine into a single
scalar gauge parameter $\varepsilon(x)$. For the spin $s=2$
case one shows that the gauge transformation of the frame with $o(1,1)$ vector parameter $\varepsilon^a(x)$
and the diffeomorphism with parameter $\xi^n(x)$ are identified \cite{Jackiw:1978ar}.
For the higher spins $s>2$  diffeomorphism parameters  form a subspace in the  gauge parameter
space.}

\subsection{Linearization around  $\ads$ background}

The  higher spin theory \eqref{action}
contains  the gravitational subsector since the higher spin algebras under consideration always contain $o(2,1)$
subalgebra. Moreover, the ground state of the model is identified with the $\ads$ spacetime.
It seems natural to have $\ads$ spacetime as the background,
because in this way  higher dimensional higher spin gauge
theories extend to the  $2d$ case while  keeping their main
characteristic features intact: higher spin gauge fields and the
$AdS$ background geometry. One should note, however, that contrary
to $d \geq 4$ higher spin theories the $\ads$ background is not necessarily
required   to have a consistent interacting theory. \footnote{See,
\textit{e.g.}, Refs. \cite{Afshar:2013vka}, where $3d$ flat higher spin theory
was discussed.} Recall that switching on the cosmological constant
$\Lambda \neq 0$ is indispensable to guarantee consistent gravitational
interactions of gauge massless higher spin fields. In two and
three dimensions it seems that taking $\Lambda = 0$ does not prevent
having a consistent theory with higher spin symmetries  because
higher spin   fields carry no local degrees of freedom.

Fixing the background connection $W_0$ we treat dynamical fields $\Omega$ as fluctuations,
\be
\label{decback}
W(Y|x) = W_0(Y|x) + \Omega(Y|x)\;,
\ee
where $W_0$ satisfies the $o(2,1)$ zero-curvature condition \eqref{zerocurv} and describes $\ads$ spacetime.
A background value of $\Psi$ is discussed below, while  perturbations over $\Psi_0$ are defined as
\be
\label{decpsi}
\Psi(Y|x) = \Psi_0(Y|x) + \Phi(Y|x)\;,
\ee
where $\Phi$ are dynamical fields. Up to the second order in the fields  the non-linear curvature \eqref{fullcurvature}
decomposes as
\be
\cR(Y|x) = \cR_0(Y|x) + R(Y|x) + ... \;,
\ee
where
\be
\cR_0 = d W_0 + W_0 * W_0\;,
\qquad
R = d \Omega + W_0 * \Omega +\Omega * W_0\;.
\ee

Substituting the perturbative expansions  \eqref{decback}, \eqref{decpsi} into the equations of motion
\eqref{covcon}, \eqref{curveq} one finds that the background fields satisfy the following equations
\be
\label{backeqs}
\ba{l}
\dps
d W_0 + W_0  *\, W_0 = 0\;,
\qquad
d \Psi_0 + [W_0,\Psi_0]_*  = 0\,.
\ea
\ee
The first equation above is the zero curvature-condition \eqref{zerocurv}, while the background
field $\Psi_0$ remains unknown. Next, the first-order equations  are given by
\be
\label{firstorder}
d \Omega + [W_0, \Omega]_ \star = 0\;,
\qquad
d \Phi + [W_0,    \Phi]_*  + [\Omega,   \Psi_0]_* = 0\,.
\ee

Suppose now that $\Psi_0$ is $x$-independent, that is $d\Psi_0 = 0$.
Then, the second equation in \eqref{backeqs} says that
\be
[W_0, \Psi_0]_* =0\;.
\ee
It follows that $o(2,1)$-invariant non-vanishing vacuum value of the $0$-form field  is a function of the
$sp(2)$ basis elements only
\be
\label{vaccumzeroform}
\Psi_0(Y) = a_{(0)}  + a_{(0)}^{\alpha\beta}t_{\alpha\beta} +a_{(1)} c_2+ ... \;=  \sum_{k,\,l=0}^\infty a_{(l)}^{\alpha_1\alpha_2 ... \alpha_{2k}}\; T_{\alpha_1\alpha_2 ... \alpha_{2k}}*(c_2*)^l
\;,
\ee
where $a_{(l)}^{\alpha_1\alpha_2 ... \alpha_{2k}}$ are some $(Y,x)$-independent (constant) $sp(2)$ symmetric
tensor parameters, $T_{\alpha_1\alpha_2 ... \alpha_{2k}}$ is given by \eqref{PPP} and $c_2$ is  $sp(2)$
Casimir operator. \footnote{Choosing  $\Psi_0 = t_{\alpha\beta} a^{\alpha\beta}$ in \eqref{vaccumzeroform} is
similar to non-vanishing vacuum value of the $0$-form in the BF higher spin  model considered in
Ref. \cite{Vasiliev:1995sv}.} Recall that these properties guarantee the $sp(2)$ invariance of $\Psi_0$,
cf. \eqref{spdecom}. The fluctuation field  $\Omega$ is also $sp(2)$ invariant,   and therefore
it commutes with any combination of $t_{\alpha\beta}$. As a result, $[\Omega, \Psi_0]_* = 0$.

It follows  that the linearized equations of motion \eqref{firstorder} take the form
\be
\label{eqsfluc}
\ba{l}
\dps
d \Omega + [W_0, \Omega]_* = 0\;,
\qquad
d \Phi + [W_0,    \Phi]_*   = 0\,.
\ea
\ee
The Abelian part of the gauge transformation \eqref{HSgaugetr} for
fluctuations has the form
\be
\label{HSgaugetr2}
\delta_\varepsilon \Omega  = D_0 \varepsilon \equiv d \varepsilon +[W_0, \varepsilon]\;,
\qquad
\delta_\varepsilon \Phi = 0\;,
\qquad
\delta_\varepsilon R =  0\;,
\ee
where the linearized derivative $D_0$ reproduces the definition \eqref{D0}, while the above transformations
themselves reproduce \eqref{transOmega} and \eqref{transPhi}.

Now, the trace decomposition \eqref{spdecom} that brings the higher spin  algebra
$hc(1|2\hspace{-1mm}:\hspace{-1mm}[1,2])$ into the basis where all
basis elements are given by traceless $o(2,1)$ tensors
\eqref{plot} is expressed via the $sp(2)$ generators. It follows that field
$\Omega_m$ decomposes into irreducible components as
\be
\label{gaugedec}
\Omega_m :=\bigoplus_{s=1}^\infty \bigoplus_{k=0}^\infty \,\Omega^{(s,k)}_m\;,
\ee
where components $\Omega^{(s,k)}_m$ are  $1$-form spin-$s$ gauge  fields
$\Omega^{(k)}{}_m^{A_1... A_{s-1}}$ with $s-1$ totally symmetric traceless $o(2,1)$ indices,
while the label $k$ stands for a multiplicity, cf. \eqref{infsln}.

On the other hand, field equations \eqref{eqsfluc} can be represented via the
background covariant derivative as  $D_0 \Omega =0$ and $D_0 \Phi
= 0$, cf. \eqref{2eq}, \eqref{1eq}.
 Therefore, using   $D_0
t_{\alpha\beta} = 0$ one finds out that the field equations
\eqref{eqsfluc} can be decomposed into $o(2,1)$ irreducible
components as well.
In each irreducible spin-$s$ sector equations
of motion take the form \eqref{linBFeom}; each pair of equations
\eqref{linBFeom} comes in infinitely many copies. Whence, the
spectrum of the model contains infinitely many copies of all
integer spin-$s$ subsystems,
\be
1_{[\infty]},\;\;2_{[\infty]},\;\; 3_{[\infty]}, \;\; ...\;\; , \infty_{[\infty]} \;,
\ee
where $1,2,3,..$ denote spins, while a subscript $[\cdot]$ denotes a multiplicity, which in the present case
is infinite, cf. \eqref{gaugedec}.

\subsection{Reduced BF higher spin models}

The spectrum of the $\ads$ higher spin gravity  model \eqref{action}  is infinite and degenerate. It can be truncated in two possible ways.

\begin{itemize}

\item Horizontally reduced model: finitely many fields with spins bounded from above, each field appears in several copies.

\item Vertically reduced model: infinitely many fields of all spins from zero to infinity, each field appears in a single copy.
\end{itemize}

It is clear that such reduced models are governed by respectively horizontal and vertical
quotient higher spin algebras of Section  \bref{sec:quotients}.

We propose to describe reduced models with fields taking values in the quotient higher spin algebras  by the
BF action \eqref{action} modified by the  projecting operator $\Delta$ in the following manner
\footnote{Action functionals of this type were previously considered within  $AdS_5$ higher spin interacting theories
\cite{Vasiliev:2001wa,Alkalaev:2002rq,Alkalaev:2010af}}
\be
\label{action_fact}
S_\Delta [\Psi, W] = g \int_{\cM^2} \tr \Big[\Delta* \Psi *  \cR \Big]\;,
\ee
where, according to particular factorization, one chooses either the horizontal projector $\Delta_k$ or
the vertical projector $\Delta_u$ of
Section \bref{sec:factor}. By inserting $\Delta$ we reduce the original spectrum of fields
to a smaller subset of fields identified with representatives of the quotient algebra. Indeed,
$\Delta$ is defined to send all elements of the corresponding ideals in $hc(1|2\hspace{-1mm}:\hspace{-1mm}[1,2])$
to zero \eqref{defpro}.

Action \eqref{action_fact} can be understood by introducing a new invariant form.
Indeed, we replace the invariant form \eqref{invform} on the algebra $hc(1|2\hspace{-1mm}:\hspace{-1mm}[1,2])$
by the following form
\be
\label{invformDelta}
\langle F, G  \rangle_{\Delta} = \mathrm{Tr}(\Delta*F *G)\;,
\qquad
F, G \in hc(1|2\hspace{-1mm}:\hspace{-1mm}[1,2])\;.
\ee
The invariance and symmetry  properties are not spoiled by  $\Delta$ as it commutes with $F$ and $G$, \eqref{defpro}. However, the invariant
form \eqref{invformDelta} is degenerate since  $\langle F, G  \rangle_{\Delta} =0$ for $\forall\, F \in hc(1|2\hspace{-1mm}:\hspace{-1mm}[1,2])$
and $\forall \,G \in \cI$.

Reduced action \eqref{action_fact} is invariant with respect to the  gauge transformations \eqref{HSgaugetr}.
Additionally, it acquires a new type of invariance due to a degeneracy of the form \eqref{invformDelta},
\be
\label{shift}
\ba{c}
\dps
\delta \Psi(Y|x) = A(Y|x)\;, \qquad A \in \cI\;,
\\
\\
\dps
\delta W(Y|x) = B(Y|x)\;, \qquad B \in \cI\;.
\ea
\ee
If the factorization with respect to the ideal $\cI$ gives a quotient algebra which is not simple,
then there happens a symmetry enhancement governed by an additional  ideal. This is the case of the
double factorization described in Sections \bref{sec:doublefack} and \bref{sec:factor}.

The equations of motion of the reduced theory  \eqref{action_fact} are
\be
\label{covconDelta}
\Delta*\cR_{mn}(Y|x)  = 0\;,
\ee
and
\be
\label{curveqDelta}
\Delta*D_m \Psi(Y|x) = 0\;,
\ee
where the  covariant derivative $D_m$ is given by \eqref{derivative}. The equations
are invariant with respect to the standard gauge transformations, while the shift transformations \eqref{shift}
yield additional algebraic Bianchi identities.

Let us consider a perturbative expansion of the reduced model
\eqref{action_fact}. Both zeroth-order and first-order equations
are again equations \eqref{backeqs} and \eqref{firstorder} but now
multiplied by $\Delta$. A natural choice for the background is to
take the $\ads$ connection  $W_0$  as the vacuum  $1$-form field because it solves
the equation of motion \eqref{covconDelta}. As the background
$0$-form field we take an $x$-independent $\Psi_0(Y)$. From
\eqref{curveqDelta} it follows  that $\Delta * [W_0, \Psi_0]_* =0$
which means that $\Psi_0$ can be chosen to be an element of the ideal, $\Psi_0 \in \cI$. However,
using the shift symmetry
\eqref{shift} one observes  that it can be equivalently set
to zero. Therefore, from the very outset one can choose  $W = W_0$
and $\Psi_0 = 0$ as representatives of the zeroth equivalence class in the quotient
higher spin algebra.

On the other hand, the projector is $o(2,1)$-invariant since $D_0
\Delta(Y) = 0$, where $D_0$ is the background $o(2,1)$ covariant
derivative \eqref{D0}. Introducing the quotient algebra
representatives
 $\bar \Omega  = \Delta * \Omega$ and $\bar \Phi  = \Delta * \Phi$ one rewrites the linearized
equations of motion as $D_0 \bar \Omega(x|Y) = 0$ and $D_0 \bar \Psi(x|Y) = 0$. It follows that
the linearized equations  factorize
into independent spin-$s$ subsystems described by previously studied equations \eqref{linBFeom}.

In the case of the horizontal factorization, the respective  quotient higher spin algebra is given by  a direct sum of
general linear  algebras
\eqref{horizontalgebra}. It follows that for a given parameter of the horizontal factorization
$k = 1,2, ...$, a spectrum of the reduced model is degenerate. It contains independent
subsystems of spins:
\be
2k_{[1]},\;\; (2k-1)_{[1]}, \;\; (2k-2)_{[2]}, \;\; (2k-3)_{[2]}, \;\; (2k-4)_{[3]}, \;\; (2k-5)_{[3]}, \;\;...
\ee
where $2k - i$ denotes spin, while a subscript $[j]$ denotes a
multiplicity. Spin-$1$  and spin-$2$ subsystems have a maximal
multiplicity  $[k]$. For instance, the maximal horizontal
factorization ($k=1$) gives spin $s =(2_{[1]}, 1_{[1]})$ system
that obviously reproduces the original Jackiw-Teitelboim model
plus the Maxwell BF theory. A spectrum of  the next-to-maximal horizontal
factorization ($k=2$) reads $4_{[1]}, 3_{[1]},
2_{[2]}, 1_{[2]}$.

In the case of the vertical factorization, the resulting higher spin algebra $\Hsalg$ is infinite-dimensional
and parameterized by continuous   parameter  $\nu$. A spectrum of the reduced model is non-degenerate. It
contains independent subsystems of spins:
\be
\nu \neq \nu_0\;:\qquad 1_{[1]},\;\; 2_{[1]}, \;\; 3_{[1]},  \; ... \; , \infty_{[1]} \;.
\ee
Generally, the spectrum does not depend on $\nu$, but for the special values \eqref{u0}
it is truncated to a finite subset of subsystems with spins:
\be
\nu_0= (k-1)(2k-1)\;: \qquad 1_{[1]},\;\; 2_{[1]}, \;\; 3_{[1]}, \;\;   ...\;, (2k-1){}_{[1]}, \;\; (2k){}_{[1]} \;,
\ee
that immediately follows from that the reduced higher spin algebra is $gl(2k, \mathbb{R})$ \eqref{feigin}.
\footnote{One can also discuss reduced models based on double factorizations of the form \eqref{Hp}.}

\section{Conclusions and outlooks}
\label{sec:concl}

In this paper, we proposed  a new class of two-dimensional higher spin models interpreted as the
$\ads$ higher spin gravity and explored some of its global and local properties. The model is formulated by virtue of
 topological  BF action  for  fields taking values in  particular higher spin symmetry algebra
containing   $o(2,1) \approx sl(2, \mathbb{R})$ subalgebra. Our analysis follows methods
used within the unfolded approach to higher spin dynamics. In particular, we developed a two-dimensional
version of the unfolded formulation  resulting in a cohomological understanding of the BF
dynamics. Using  two different nilpotent operators acting  on the field space of BF model we
elaborate two metric-like formulations of the model. Our analysis of the linearized BF
equations of motion both for $0$-forms and $1$-forms accomplishes the analysis of the
1-form sector performed earlier in \cite{Alkalaev:2013fsa}. We also discuss a new type of duality
between two metric-like formulations obtained from a single BF frame-like theory.

We suggested a particular formulation of  two-dimensional  higher spin algebra $\Hsalg$  employing
the $o(2,1)-sp(2)$ Howe duality. In this way we extend the Vasiliev oscillator construction of $d\geq 4$
higher spin Eastwood-Vasiliev algebras to the $d=2$ case. Infinite-dimensional higher spin algebras and their finite-dimensional truncations
are realized  as particular quotient algebras for which reason we classified relevant cases of ideals and
corresponding factorizations. We explicitly described the projecting technique used to define the BF actions for fields taking values in the quotient
algebras.

The $d=2$ classification of ideals and factorizations extends to
any $d$ case. Obviously, using the ideals generated by the $sp(2)$
Casimir operator and its powers one arrives at some  quotient
algebra with connections identified with higher spin
partially-massless fields of any depth (\textit{e.g.}, see  discussion in \cite{Alkalaev:2007bq}). It should be realized as
the symmetry algebra of higher order singleton representations of
$o(2,d)$ algebra \cite{Bekaert:2013zya}.

It is important to note that a given BF theory with a
finite-dimensional algebra is necessarily topological one. The
situation is more intricate in the  case of an
infinite-dimensional algebra. For instance, the BF action for
higher spin algebras considered in this paper  is topological. On
the other hand, a particular  BF theory  proposed in Ref.
\cite{Vasiliev:1995sv} describes self-interactions of matter
fields via higher spin currents built of these matter fields.
Nonetheless, the model is not topological because BF fields  take
values in a peculiar  infinite-dimensional algebra containing
$\Hsalg$ as a subalgebra. The rationale
behind this  observation is that a BF action
formulated on an infinite-dimensional field space may leave a room
for local degrees of freedom.

In particular, it follows that BF actions may contain current
interactions of matters fields, and, therefore, it is tempting to
speculate that higher spin BF action has to do somehow both
with currents and matter fields on equal footing. This idea
conforms with the duality between  the metric-like formulations
described  in this paper. Indeed, we find out that BF
equations of motion can be simultaneously treated as matter field
equations and conservation conditions.

Below we list some interesting issues left beyond the scope of the paper.

\begin{itemize}

\item The form and properties of the mapping between two
metric-like descriptions of the free field  higher spin theory
discussed in Section \bref{sec:mapping}.
The original linearized  BF higher spin action functional
can be treated as a parent action for the two dual formulations.

\item One may consider the supersymmetric  Howe dual pair $o(2,M)
- osp(1,2)$ underlying the construction of the higher spin algebra
$hc(1|(1,2)\hspace{-1mm}:\hspace{-1mm}[M,2])$ which describes
hook-type  mixed-symmetry higher spin fields in $AdS_{M+1}$
\cite{Vasiliev:2004cm}. For $M=1$ all mixed-symmetry fields are
dual to totally symmetric ones \eqref{Hodge}. One can classify ideals of
$hc(1|(1,2)\hspace{-1mm}:\hspace{-1mm}[1,2])$ as in Section
\bref{sec:quotients}, and study respective quotient algebras. In
particular, it should result in odd values of the rank of  general
linear algebras obtained via the double factorization
\eqref{feigin}.

\item It is interesting to realize the universal enveloping
algebra $\cU(o(2,1))$ in terms of extended  $
o(2,1)-osp(n,2)$ Howe dual pairs with arbitrary $n \geq 2$.

\item The role of parameter $\nu$ in the vertical reduced model is
to be clarified. We have seen that the linearized equations of
motion are independent on $\nu$.  It appears that $\nu$ comes out  in the
next orders. \footnote{See recent paper \cite{Boulanger:2013naa} on $3d$ Chern-Simons
higher spin theories, where the parameter has been related to a
spin of infinite-dimensional anyon representations in $AdS_3$.}

\item The flat space limit $\Lambda \rightarrow 0$ in
the BF higher spin models. The resulting theory should be a
higher spin extension of the two-dimensional Poincare gravity suggested in
\cite{Callan:1992rs} and further discussed in
\cite{Jackiw:1992bw,Cangemi:1992bj}. It should be governed by a
non-semisimple higher spin algebra extending  the $(1+1)$ Poincare
algebra.

\end{itemize}

Among other things, the $\ads$ higher spin gravity is interesting because
the respective action functional is given in a closed
form that makes possible to analyze many conventional questions like higher spin black hole solutions,
supersymmetric higher spin extensions, quantization, etc. In particular, it is interesting to
consider matter fermions interacting via higher spin  fields and,
therefore, to formulate a higher spin extension of the Schwinger
model in $\ads$ spacetime. \footnote{\textit{E.g.,} see a
discussion of a particle moving in lineal gravitational fields
\cite{Cangemi:1992bj}.} Further, topological field theories are
known to induce  local  degrees of freedom at the boundary. This is
also the case for two-dimensional higher spin theories of the type
considered in the present paper. The problem has been already
partly  discussed in the literature \cite{Rey,Grumiller:2013swa}.

\vspace{5mm}

\noindent \textbf{Acknowledgements.} I am grateful to M.A.
Grigoriev, S.E. Konstein, R.R. Metsaev, and E.D. Skvortsov for
many  useful discussions and comments.

\noindent This research was supported by Russian Science Foundation grant 14-42-00047.

\appendix

\section{Computation of the cohomology groups }
\label{sec:appendixA}

In what follows, we compute the cohomology of the nilpotent $\sigma_{\pm}$ operators acting
on the space $\cG_s$. To this end, one recalls
some relevant group-theoretical facts on $o(1,1)$ Lorentz algebra
representations and their tensor products.

Introducing a collective notation for symmetrized indices $(a_1...a_k)
\equiv a(k)$, one finds that a frame-like tensor
$\text{T}_m{}^{a(k)}$ being  a tensor product of totally symmetric
and traceless tensor with a vector  decomposes into two $o(1,1)$
irreps of spins $k-1$ and $k+1$. Recalling that a dimension of any
integer spin $o(1,1)$ (non-scalar) irrep equals $2$, the above
statement can be simply understood as $2^2 = 2+2$. On the other
hand, any totally symmetric and traceful frame-like tensor
$\text{A}_m{}^{a(k)}$ decomposes into $\bigoplus_{n=0}^k  \,
\text{T}_m{}^{a(n)}$, where $\text{T}_m{}^{a(n)}$ are traceless
with respect to fiber  $o(1,1)$ tensors. The decompositions
clarify the formula $\dim \text{A}_m{}^{a(k)} = 2(2k+1)$.

To summarize, the following decompositions are useful in practice
\be
\label{dectr}
\text{A}_m{}^{a(k)} = \text{A}^{a(k+1)}\oplus \text{A}^{a(k-1)}\;,
\ee
\be
\label{dectrl}
\text{T}_m{}^{a(k)} = \text{T}^{a(k+1)}\oplus \text{T}^{a(k-1)}\;,
\ee
both for traceful $\text{A}^{a(k)}$ and traceless $\text{T}^{a(k)}$ totally symmetric tensors.
Decomposition \eqref{dectrl} for traceless tensors is easily explained in components: a trace part is proportional to antisymmetric dualized part of
hook component. The case $k=1$ is special: decomposing
$\text{A}_m{}^{a} \equiv \text{T}_m{}^{a}$ into $sl(2)$ irreps and then into $o(1,1)$ irreps yields
\be
\label{decspin2}
\text{A}_m{}^{a} \equiv \text{T}_m{}^{a} := \text{A}^{a(2)}\oplus \text{A}
= \text{T}^{a(2)}\oplus  \text{T} \oplus \text{A}\;,
\ee
where $A$ and $T$ are two different scalar components. Their appearance is due to the  relation
$A^{a|b} = \half \text{A}^{(a|b)} + \half \text{A}^{[a|b]} = \half \text{A}^{(a|b)} + \half \epsilon^{ab} \text{A} =
\half \text{T}^{(ab)} + \frac{1}{4} \eta^{ab}\text{T} + \half \epsilon^{ab} \text{A}$,
where $\eta_{mn}\text{T}^{(mn)}=0$ and $\epsilon^{ab}$ is $2d$ Levi-Civita tensor. Vertical slash denotes
independent groups of indices.

Consider operators $\sigma_\pm$ given by \eqref{sigmas} that act
on the module $\cG_s$ of  differential $p$-forms which take values
in $o(1,1)$ finite-dimensional irreps, $T_{(p)}^{a_1...a_k}$,
where $p=0,1,2$ and $k=0,1, ..., s-1$, see Section
\bref{sec:sigmas}. For the case $s=1$ the cohomology computation is trivial so we
give detailed consideration of the spin $s\geq 2$ case only.

\vspace{-3mm}

\paragraph{$\sigma_-$- cohomology.}
Let us compute
cohomology group $H^{(0)}(\sigma_-)$. Since exact forms are absent in this case the cohomology is defined
by the closure condition only
\be
\label{H0-}
h_c\, T_{(0)}^{a(k-1)c} = 0\;,
\qquad
0 \leq k  \leq s-1\;.
\ee
Using the background $1$-form frame  $h_{m,\, c}$
the world index is converted into fiber one so that equation
\eqref{H0-} is  cast into the form $T^{a(k-1)c} = 0$ for $k = 1,2,..., s-1$. The
case $k=0$ is exceptional: equation \eqref{H0-} does not impose any restrictions on $T$.
Thus, the cohomology group contains a single scalar component $T$,
\textit{i.e.} we find $H^{(0)}(\sigma_-) = \{T\}$, see \eqref{cohprop}.

Consider now cohomology group $ H^{(1)}(\sigma_-)$ which is defined by both closer and exactness conditions
\be
\label{H1-}
h_c \wedge T_{(1)}^{a(k-1)c} = 0\;, \qquad \delta  T_{(1)}^{a(k)}  = h_c \, T_{(0)}^{a(k)c}\;,
\ee
where $T_{(1)}^{a(k)}$ and $T_{(0)}^{a(k+1)}$,  $0 \leq k  \leq s-1$, are
$1$-forms and $0$-forms, respectively. Consider the first equation
in \eqref{H1-}. Converting all world indices into fiber ones the equation
can equivalently be rewritten as $T^{a(k-1)[c|d]} = 0$.
Contracting with $\epsilon_{cd}$ and using decomposition \eqref{dectrl} one finds  that
rank-$(k-1)$ totally symmetric and traceless component of $T_{(1)}^{a(k)}$ vanishes except
for the cases $k=0$ and $k= s-1$. Then, one considers the exactness condition in \eqref{H1-}
and shows that  rank-$(k+1)$ totally symmetric and traceless component of $T_{(1)}^{a(k)}$
also vanish since it is exact, except for the case  $k= s-1$.

Equation \eqref{H1-} at $k=1$ should be analyzed separately because in this case decomposition into
irreducible components is different, see \eqref{decspin2}. It follows that the closer condition
sets to zero the antisymmetric part, while symmetric one is arbitrary. For $s>2$ symmetric and
traceless component cancels due to the exactness condition, while for $s=2$ it remains intact.
One concludes that cohomology is given by rank-$s$ totally symmetric component and
a scalar component $T$ which comes as a trace part of $T^a_{(1)}$. Therefore,
$H^{(1)}(\sigma_-) = \{T, T^{a_1 ... a_{s}}\}$, see \eqref{cohprop}.

Then, consider  cohomology group $ H^{(2)}(\sigma_-)$  defined by the following chain of conditions
\be
\label{H2-}
h_c \wedge T_{(2)}^{a(k-1)c} \equiv  0\;,
\qquad
\delta T_{(2)}^{a(k)}  = h_c \wedge T_{(1)}^{a(k)c}\;,
\qquad
\delta T_{(1)}^{a(k)}  = h_c \, T_{(0)}^{a(k)c}\;,
\ee
where $T_{(2)}^{a(k)}$, $T_{(1)}^{a(k+1)}$, and $T_{(0)}^{a(k+2)}$,  $0 \leq k  \leq s-1$, are respectively
$2$-forms, $1$-forms, and $0$-forms. Being a $3$-from the first equation in \eqref{H2-} is identically
satisfied. On the other hand, analysis of the exactness conditions in \eqref{H2-} is similar
to previously done  computation of  $ H^{(0)}(\sigma_-)$  and $ H^{(1)}(\sigma_-)$.
Repeating the reasoning we find that $H^{(2)}(\sigma_-) = \{T^{a_1 ... a_{s-1}}\}$, see \eqref{cohprop}.

\vspace{-3mm}

\paragraph{$\sigma_+$- cohomology.}  Computation of  $\sigma_+$ cohomology is analogous.
The only essential difference is the origin of the scalar component in $H^{(1)}(\sigma_\pm)$:
for the case of $\sigma_+$  this is an antisymmetric
component of $A^{m|n}$, while for the case of $\sigma_-$  the scalar component is identified with the trace
of $A^{m|n}$, cf. \eqref{decspin2}. The resulting cohomology groups $H^{(p)}(\sigma_+)$ are given
in \eqref{cohprop}.

\section{Horizontal and vertical (quasi-)projectors}
\label{sec:appendixB}

\vspace{-2mm}
\paragraph{Horizontal  projection.} Substituting \eqref{ideal1} into \eqref{defpro}
one gets a  function $\Delta_k(z)$ satisfying the horizontal projecting  equation
\be
\label{mastereq1}
\Delta_k* T_{\alpha_1 ... \alpha_{2k}} = \big[\, D^{(k)} \Delta \,\big] T_{\alpha_1 ... \alpha_{2k}} = 0\;,
\ee
where $D^{(k)}$ stands for $k$-th degree of the second-order differential operator
\be
\label{operD}
D = 2z \frac{d^2}{ d z^2} + 2 \frac{d}{ d z} +1\;.
\ee
The ordinary differential equation $D^{(k)} \Delta_k  =0$    has $2k$ independent solutions.
Among them we single out only those that have the form of the series
$\Delta = \kappa_0z^\alpha + \kappa_1 z^{\alpha+1}+ \kappa_2 z^{\alpha+2} + \cdots$, for some $\alpha \geq 0$. It turns
out that $\alpha = 0$ and there are $k$ independent solutions of this type, $\Delta_i$, $i=1,...,k$. Since equation $D^{(k)} \Delta  =0$
comes as differential consequences of equation $D^{(k-1)} \Delta  =0$, one concludes that $k-1$ solutions
$\Delta_i$, where $i = 1,...,(k-1)$ solve equation of lower rank and therefore can be found by induction, while the
highest rank solution $\Delta_k$ does describe factorization \eqref{mastereq1}. From the algebraic perspective,
a set of analytical solutions to the horizontal projecting equation  is clearly explained by
the flag sequence of ideals \eqref{flag}.

An explicit form of solutions can be found straightforwardly  provided that differential
operator \eqref{operD} is represented as $\dps D = 2(N_z
+1)\,\frac{d}{dz} +1$, where $\dps N_z = z \frac{d}{dz}$ is the Euler operator, so that searching for a solution in the form of
power series yields a recurrent equation system.

Solutions to equation \eqref{mastereq1}
can be expressed via the  Bessel functions and their multiple
integrals. For instance, in the case $k=1$  equation \eqref{mastereq1}
is in fact  the Bessel equation of zeroth order solved by
\footnote{In $d$ dimensions
the $k=1$ equation  describes the maximal factorization; the solution is given in the particular
integral form \cite{Vasiliev:2004cm}.}
\be
\label{bessel}
\Delta_{k=1}(z) = I_0(\sqrt{2z})\;.
\ee
In the case $k\geq 2$ equation \eqref{mastereq1} can
be expressed via auxiliary combinations $F_m(z) = D^{(k-m-1)}
\Delta(z)$ as inhomogeneous Bessel equation $D F_m (z) =
F_{m-1}(z)$, where $m = 0, ...,k-1$ and $F_{k-1}\equiv \Delta$.

It is worth noting that using the horizontal factorization via projector \eqref{mastereq1}  yields
finite-dimensional quotient algebras \eqref{horizontalgebra} with basis elements realized as infinite
formal power series of auxiliary variables $Y^A_\alpha$, and not as bilinear combinations as
one might expect from \eqref{basis}.

\vspace{-4mm}

\paragraph{Vertical projection.} Substituting \eqref{ideal2} into \eqref{defpro}
one gets  a function $\Delta_\nu(z)$ satisfying the vertical projecting condition expressed as the
$4$-th order differential equation
\be
\label{mastereq2}
\Delta_\nu*(c_2 + \nu) = z^2 F^{\prime\prime} + 4z F^\prime +\frac{1}{2} z F+ \frac{9}{4} F + \nu \Delta_\nu = 0\;,
\qquad
F = D \Delta_\nu\;,
\ee
where differential operator $D$ is given by \eqref{operD}.
Solutions analytical in $z=0$ have  the form
$\Delta_\nu(z) = \gamma_0 + \gamma_1 z+ \gamma_2 z^{2} + \cdots$, where the coefficients
satisfy the following recurrent equation system
\be
\label{rec1}
9\gamma_1 + \big(2\nu+\frac{9}{2}\big) \gamma_0 = 0\,,\;\; \gamma_{k-2}+ A_k \gamma_k + B_k \gamma_{k-1} = 0\;,
\ee
where  $A_k$ and $B_k$  are given by
\be
\label{rec2}
A_k = k^2(2k+1)^2\;,
\qquad
B_k = 2(k-1)(2k+1) +2\nu+\frac{9}{2}\;.
\ee
A few first coefficients for $\gamma_0 = 1$ are found to be
\be
\label{solution}
\Delta_\nu(z)  = 1 - \frac{u_\nu}{3^2}\,z +\frac{u_\nu(10+u_\nu)-9}{(30)^2}\, z^2  + \cdots\;,
\quad \text{where}\quad u_{\nu} = 2\nu+9/2\;.
\ee

Following the discussion of the double factorization in Section \bref{sec:doublefack},
one observes that given  a particular  value \eqref{u0}
quotient $\cH_{\nu_0}$ defined by projecting condition
\eqref{mastereq2} possesses an additional ideal formed by elements
proportional to \eqref{PPP}. Indeed, using relation \eqref{P2} one
shows  that operator  $\Delta_{\nu_0}$  satisfying  the
projecting condition $\Delta_{\nu_0} * (c_2 + \nu_0) = 0$ can be
represented in the form
\be
\label{Projrelation}
\Delta_{\nu_0} =  \Delta_k*\,\prod_{m=0}^{k-2} *\,(c_2 + \alpha_m)\;,
\ee
where $\Delta_k$ fulfills the horizontal projecting condition \eqref{mastereq1}. It follows that elements of the quotient
$\cH_{\nu_0}$ proportional to \eqref{PPP} are sent to zero  by
virtue of the projecting property of the prefactor  $\Delta_k$.

For instance, taking $k=1$ corresponding to $\nu_0 = 0$ \eqref{u0} one finds from \eqref{Projrelation}
that the vertical and horizontal  projectors coincide, $\Delta_{\nu_0 = 0} =  \Delta_{k=1}$. In particular,
substituting $\nu_0 =0$ into \eqref{rec1}-\eqref{rec2} one finds the  solution \eqref{solution} in a closed form
$\dps\Delta_{\nu_0 = 0}(z) = \sum_{k=0}^\infty \frac{(-)^k}{2^k(k!)^2}z^k$ recognized as the Bessel function,
$\Delta_{\nu_0 = 0}(z) = I_0(\sqrt{2z})$ \eqref{bessel}. On the other hand, we know that the $k=1$ horizontal projection
yields the quotient  $\cH_k \approx  gl(2, \mathbb{R}) $ \eqref{horizontalgebra}, while the double factorization
in the case $\nu_0=0$ yields $\cH^1_{\nu_0}/\cI_1 \approx  gl(2, \mathbb{R})$ \eqref{feigin}. The resulting quotients
obviously coincide. Note, however, that for $k>1$  the horizontal quotient algebra $\cH_k$ and the double
quotient algebra $\cH^1_{\nu_0}/\cI_k$
are not isomorphic anymore, while the respective projectors do not
coincide as well, see \eqref{Projrelation}.

\vspace{1cm}


\addtolength{\baselineskip}{-3pt}
\addtolength{\parskip}{-3pt}
\providecommand{\href}[2]{#2}\begingroup\raggedright\endgroup

\end{document}